\documentclass[aps,twocolumn,amsfonts,amssymb,pra,superscriptaddress]{revtex4}
\usepackage{graphicx}
\usepackage{epstopdf} 
\usepackage{dcolumn}
\usepackage{bm}
\usepackage{bbm,bbold}
\usepackage{color}
\usepackage{xcolor, soul}
\sethlcolor{green}
\usepackage{amsfonts}
\usepackage{amsmath}
\usepackage{mdframed}
\usepackage[normalem]{ulem}
\usepackage{mathrsfs}   
\usepackage[none]{hyphenat}
\usepackage{subfigure}
\usepackage{float}
\usepackage[colorlinks=true,citecolor=blue]{hyperref}
\hypersetup{colorlinks=true,citecolor=blue,linkcolor=blue,urlcolor=blue}

\DeclareMathAlphabet\mathbfcal{OMS}{cmsy}{b}{n}

\allowdisplaybreaks 

\begin{document}

\title{Loss-induced collective mode in one-dimensional Bose gases} 

\author{Jeff Maki}
\affiliation{Pitaevskii BEC Center, CNR-INO and Dipartimento di Fisica, Universit\`a di Trento, I-38123 Trento, Italy}

\author{Lorenzo Rosso}
\affiliation{Universit\'e Paris-Saclay, CNRS, LPTMS, 91405, Orsay, France}

\author{Leonardo Mazza}
\affiliation{Universit\'e Paris-Saclay, CNRS, LPTMS, 91405, Orsay, France}

\author{Alberto Biella}
\affiliation{Pitaevskii BEC Center, CNR-INO and Dipartimento di Fisica, Universit\`a di Trento, I-38123 Trento, Italy}

\date{\today}

\begin{abstract}
 We show that two-body losses induce a collective excitation in a harmonically trapped one-dimensional Bose gas, even in the absence of a quench in the trap or any other external perturbation. Focusing on the dissipatively fermionized regime, we perform an exact mode expansion of the rapidity distribution function and characterize the emergence of the collective motion. We find clear coherent oscillations in the widths of the gas both in position and rapidity space as well as in the phase space quadrupole mode of the gas.  We compare this motion to that of the well known breathing mode in the absence of losses and find that they differ in the motion of the density profile.
 We also discuss the role of losses in the collective motion of the gas following various other protocols, such as the removal of losses after a finite hold time, and when the harmonic trap is also quenched.
\end{abstract}
\maketitle

\section{Introduction}
\label{sec:introduction}
Collective modes in harmonically trapped atomic gasses are long-lived excitations involving the coherent motion of the system as a whole.
The dynamics of such excitations contain important information
about quantum correlations, the nature of the interactions, and the symmetries of a many-body system. 
This class of modes has been the subject of extensive theoretical studies and many such excitations have been observed experimentally \cite{Menotti02,Minguzzi05, Fang14, pitaevskii2016bose}.

This physics is usually studied in settings where the particle number and energy are conserved quantities, that is when the system's dynamics are ruled by unitary time evolution. In order to excite these modes, it is necessary to abruptly change the harmonic trapping potential or induce some other small density perturbation. Similarly, the sudden coupling of a system to an environment of any form will leave the system in a non-equilibrium state.
It is an open question whether the subsequent redistribution of particles, momenta, and energy following a loss event can result in long-lived collective oscillations \cite{Yamamoto21} of harmonically trapped gases. 

\begin{figure}
    \centering
    \includegraphics[width=1\columnwidth]{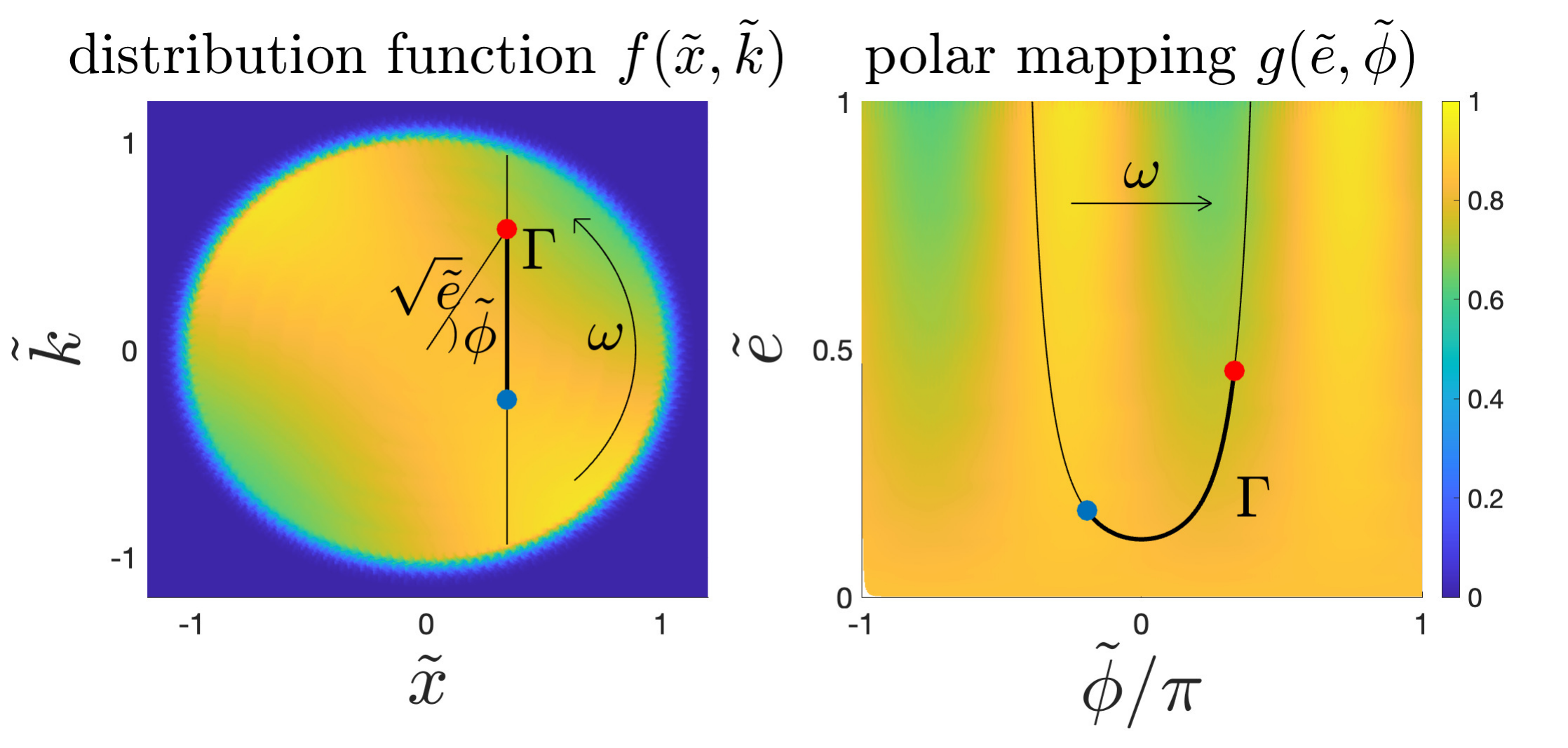}
    \caption{Typical snapshot of the rapidity distribution function at time $\tilde{t}>0$ in phase space for both planar position-rapidity $\tilde{x}-\tilde{k}$ (left panel) and  polar energy-angle $\tilde{e}-\tilde{\phi}$ (right panel) coordinates (in this sketch we set $\tilde\Gamma/\omega=1$). The harmonic confinement induces a rotation at a frequency $\omega$ (solid arrow), while losses that involve pairs of particles (blue and red dots) at the same position (black solid line) and different momenta occur at a rate $\Gamma$. Starting from an isotropic distribution function, the losses produce periodic anisotropies that oscillate at a frequency $2\omega$ as highlighted by polar coordinates. These can be detected in the moments of the distribution function.} 
    \label{fig:sketch}
\end{figure}

In this article we study a loss induced collective mode (LICM) in a paradigmatic example of an open quantum system: the Lieb-Liniger model of
one-dimensional (1D) interacting bosons with two-body losses \cite{Syassen08,Duerr09,Zhu14, Johnson_2017,Tomita17, Bouchoule20,Bouchoule20b,Bouchoule21, Rossini21,Rosso22, Rosso22b, Rosso23}. Such losses can be readily achieved in experiment using photoassociation \cite{Tomita17} or inelastic two-body scattering \cite{liu2022weakly}.  
In the limit of strong two-body  losses or strong local interactions the gas becomes dissipatively fermionized \cite{Zhu14, Syassen08,Duerr09, Rossini21,Rosso22, Rosso22b, Rosso23}; states with multiple bosons at a given position are quickly removed after several loss events, while fermionized (hard-core) bosonic states become weakly dissipative.
The consequences of these fermionic correlations on the number of particles has been examined both in free-space \cite{Syassen08,Duerr09, Rossini21} and in harmonic traps \cite{Rosso22,Gerbino23, Perfetto23, Riggio23}. 

Using a lossy Boltzmann equation for the density distribution and performing its exact mode expansion in phase space, we identify and characterize signatures of the LICM in the oscillations of the widths of the gas in both position and rapidity space at exactly twice the harmonic trap frequency.
Our analysis shows that this collective mode at leading order is a quadrupolar oscillation in phase space.
Remarkably, such mechanism is uniquely enabled by losses and persists up to all times accessible with our numerics.

The remainder of this article is summarized as follows. In Sec.~\ref{sec:BE} we put forward a Boltzmann equation to capture the dynamics of a lossy 1D Bose gas in the fermionized regime. In Sec.~\ref{sec:dynamics} we examine the results for a lossy 1D Bose gas in the absence of quench. We then present results for two other experimental protocols in Sec.~\ref{sec:additional}. Finally we conclude in Sec.~\ref{sec:conc}. 

\section{The Boltzmann Equation and the Mode Expansion}
\label{sec:BE}
In this work we consider the Lieb-Liniger model of interacting bosons with two-body losses \cite{Rosso22}. In the presence of a harmonic trap of frequency $\omega$ the Hamiltonian reads as:
\begin{align}
    H = \int dx \psi^{\dagger}(x) &\left(-\frac{1}{2}\partial_x^2 + \frac{1}{2}\omega^2 x^2\right) \psi(x) \nonumber \\
    &+ \frac{g}{2} \int dx \psi^{\dagger}(x)\psi^{\dagger}(x) \psi(x)\psi(x),
\end{align}
where $\psi(x)$ is a bosonic annihilation operator, $g$ is the elastic two-body interaction strength. In this definition and hereafter
we set $\hbar=m=1$ where $m$ is the particle mass. In the presence of two-body losses, the density matrix of the system evolves according to the Lindblad master equation:
\begin{align}
    \partial_t \rho(t) &= -i  \left[H,\rho(t)\right] \nonumber \\
    &+ \gamma \int dx \left[L(x)  \rho(t) L^{\dagger}(x) -\frac{1}{2}\left\lbrace L^{\dagger}(x)L(x) ,\rho(t) \right\rbrace\right].
    \label{eq:Lindblad}
\end{align}
with $L(x) = \psi(x)\psi(x)$ is the two-body loss operator, and $\gamma$ the two-body loss rate.

The strength of the non-linear term which accounts for the two-body elastic and inelastic
interaction due to losses, can be conveniently expressed in terms of a single parameter: $\tilde{\gamma}(t) = |g-i\gamma|/n(t)$ where $n(t)$ is the instantaneous density of the system.
For strong interactions, $\tilde{\gamma}(t) \gg 1$, the system
undergoes fermionization and
the dynamics 
can be captured using a Schrieffer-Wolf transformation to derive an effective Lindblad master equation for hard-core bosons sector of the model \cite{Syassen08,Duerr09} and exploiting the exact mapping  to free fermionic modes (labelled by their rapidities $k$). These modes are effectively weakly dissipative with an effective two-body loss rate: $\Gamma = 2\gamma/(g^2+\gamma^2/4)$ that depends on both the two-body elastic, $g$, and inelastic, $\gamma$, collision strengths of the bosons. 
We note that since the particle decreases due to the two-body losses, $\tilde{\gamma}(t)$ increases with time. A natural consequence is that the description of the fermionic modes is robust at describing the long time limit, with perturbations due to the finite interaction strength becoming less important with time. 
These facts allow to exploit a 
time-dependent generalized Gibbs ensemble ansatz where the charges are the occupation of the fermionic rapidities that slowly varies in time due to $\Gamma$ giving rise to a set rate equations~\cite{Rosso22}.

In the continuum one can deduce an effective Boltzmann equation using a the local density approximation to account for the effects of the external trapping potential~\cite{Rossini21}.
Equivalently this result can be obtained by perturbatively evaluating the quantum kinetic equation in Keldysh theory at lowest order in a derivative expansion \cite{Gerbino23, Perfetto23,Huang23}:
\begin{equation}
\label{eq:Boltzmann}
\left[\partial_t + k \partial_x - \omega^2 x \partial_k \right] f_k(x,t) =- \Gamma \int_q (k-q)^2 f_q(x,t) f_k(x,t).
\end{equation}

In Eq.~(\ref{eq:Boltzmann}) we define $f_k(x,t)$ (with $x$ and $t$ being the position and the time, respectively) as the local distribution of fermionic rapidities with rapidity $k$, and $\omega$ as the trap frequency. We also define the shorthand: $\int_{q} = \int dq/(2\pi)$ and $\int_x = \int dx$ \footnote{Our definition for $\Gamma$ differs from that in Ref.~\cite{Rosso22} by a factor of $2\pi$ to match standard conventions.}.
For simplicity we will work with the units:
\begin{align}
    \tilde{x} &= \frac{x}{R} , & \tilde{k} &= \frac{ka^2}{R} , &\tilde{\Gamma} &= \Gamma \left(\frac{R}{a^2}\right)^3 , & \tilde{t} &= \omega t ,
\end{align}
\noindent where $a = 1/\sqrt{\omega}$ is the harmonic oscillator length, and $R = \sqrt{2\mu/\omega^2}$ is the Thomas-Fermi radius. The above units differ from Ref.~\cite{Rosso22} as we define $\tilde{t}$ in terms of $\omega$ to highlight the oscillatory physics.

For simplicity, we will always initialize the  distribution of the rapidities to be the zero-temperature equilibrium distribution: 
\begin{equation}
f_{\tilde{k}}(\tilde{x},0) = \theta(1-\tilde{k}^2 - \tilde{x}^2).
\label{eq:initial_state}
\end{equation}
From there, we then consider the dynamics following various protocols in Sec.~\ref{sec:additional}.


Given the distribution function at time $\tilde{t}$, it is straightforward to calculate the dynamics of physical quantities. The first is the total number of particles which is simply the integral of the distribution function over phase space \footnote{Here the integrals are defined as $\int_{\tilde{x}} = \int d\tilde{x}$ and $\int_{\tilde{k}} = \int d\tilde{k}/(2\pi)$.}:
\begin{equation}
    \frac{N(\tilde{t})}{N(0)} = 2 \int_{\tilde{x},\tilde{k}} f_{\tilde{k}}(\tilde{x},\tilde{t}),
\end{equation}
where we have normalized the number of particles at time $\tilde{t}$ to its initial value. The dynamics of the total number have been studied previously \cite{Rosso22, Gerbino23}. The dynamics of a general observables $O(\tilde{x},\tilde{k})$ at time $\tilde{t}$ is given by:
\begin{equation}
    \langle O(\tilde{x},\tilde{k}) \rangle(\tilde{t}) = \int_{\tilde{x},\tilde{k}} O(\tilde{x},\tilde{k}) f_{\tilde{k}}(\tilde{x},\tilde{t}).
\end{equation}
In what follows we will be primarily interested in the operators $O(\tilde{x},\tilde{k}) = \tilde{x}^2$ and  $O(\tilde{x},\tilde{k})=\tilde{k}^2$. These operators correspond to the width (squared) of the Fermi distribution in position and rapidity space. In terms of the original hardcore bosons, this corresponds to the potential energy in a trap with unit frequency, and the energy of the gas in the absence of the harmonic trap. For simplicity we will refer to these quantities as the potential and kinetic energies.

Instead of working with position and rapidity, it will prove to be beneficial to work with modified polar coordinates $\tilde{e} = \tilde{x}^2 + \tilde{k}^2$ and $\tilde{\phi} = \arctan(\tilde{k}/\tilde{x})$, as shown in the right panel of Fig.~\eqref{fig:sketch}. After transforming to the new basis $f_{\tilde{k}}(\tilde{x},\tilde{t}) \rightarrow g(\tilde{e},\tilde{\phi},\tilde{t})$, we expand the distribution function in terms of angular harmonics in phase space:

\begin{eqnarray}
g(\tilde{e},\tilde{\phi},\tilde{t}) &=& g_0(\tilde{e}, \tilde{t})  \nonumber \\ 
&+&\sum_{m\neq0}^{\infty} g_m(\tilde{e},\tilde{t}) \cos(2m \tilde{\phi}) + g_{-m}(\tilde{e},\tilde{t}) \sin(2m\tilde{\phi}). \nonumber \\
\end{eqnarray}

Here we consider only even harmonics as the Boltzmann equation is invariant under a parity transformation ($\tilde{k} \to -\tilde{k}$ and $\tilde{x} \to -\tilde{x}$), and our initial conditions are even under that parity transformation. In terms of these harmonics the Boltzmann equation has a simple form:
\begin{align}
    \partial_{\tilde{t}} g_m(\tilde{e},\tilde{t}) +2m g_{-m}(\tilde{e},\tilde{t}) = -\frac{\tilde{\Gamma}}{\omega}\mathcal{A}_m\left[g(\tilde{e},\tilde{\phi},\tilde{t})\right],
\label{eq:Boltzmann_mode_expansion}
\end{align}
where $m \in \mathbb{Z}$ and we use the initial condition: $g(\tilde{e},\tilde{\phi},0) = g_0(\tilde{e},0)=  \theta(1-\tilde{e})$.
The left hand side of Eq.~(\ref{eq:Boltzmann_mode_expansion}) states that modes with different $|m|$ are decoupled and that they oscillate in time $t$ at a frequency $2|m| \omega$. The coupling between modes with different $|m|$ is due to the two-body losses and is contained in the function $\mathcal{A}_m[g(\tilde{e},\tilde{\phi},\tilde{t})]$. This function depends on the specific mode $m$, and the full distribution function $g(\tilde{e},\tilde{\phi},t)$:
\begin{widetext}
\begin{align}
    \mathcal{A}_{m\geq0}&= (2-\delta_{|m|,0}) \int_0^1 d\tilde{e}' \int_{-min[\sqrt{\tilde{e}},\sqrt{\tilde{e}'}]}^{min[\sqrt{\tilde{e}},\sqrt{\tilde{e}'}]} dx
    \left[ \cos(2|m|\tilde{\phi}) \times\frac{1}{\pi^2}\left(\sqrt{\frac{\tilde{e}-\tilde{x}^2}{\tilde{e}'-\tilde{x}^2}} +\sqrt{\frac{\tilde{e}'-\tilde{x}^2}{\tilde{e}-\tilde{x}^2}} \right) \tilde{g}(\tilde{e},\tilde{x},\tilde{t})\tilde{g}(\tilde{e}',\tilde{x},\tilde{t})\right], \nonumber \\
        \mathcal{A}_{m<0}&= 2 \int_0^1 d\tilde{e}' \int_{-min[\sqrt{\tilde{e}},\sqrt{\tilde{e}'}]}^{min[\sqrt{\tilde{e}},\sqrt{\tilde{e}'}]} dx
    \left[ \sin(2|m|\tilde{\phi}) \times\frac{1}{\pi^2}\left(\sqrt{\frac{\tilde{e}-\tilde{x}^2}{\tilde{e}'-\tilde{x}^2}} +\sqrt{\frac{\tilde{e}'-\tilde{x}^2}{\tilde{e}-\tilde{x}^2}} \right) \tilde{g}(\tilde{e},\tilde{x},\tilde{t})\tilde{g}(\tilde{e}',\tilde{x},\tilde{t})\right].
    \label{eq:Am}
\end{align}
\end{widetext}
Note in Eq.~(\ref{eq:Am}) we have made a further change of basis by writing $\cos(\phi) = \tilde{x}/\sqrt{\tilde{e}}$ and work with a distribution function $\tilde{g}(\tilde{e},\tilde{x},t)$ which is a mix of Cartesian and polar coordinates to highlight the local nature of the two-body-loss mechanism. An in-depth derivation of Eq.~\eqref{eq:Am} is presented in Appendix \ref{app:mode_expansion}.

The value of the mode expansion is two-fold. First, relevant physics observables can be easily expressed in terms of a few harmonics:
\begin{align}
    N(\tilde{t}) &= \int_0^\infty d\tilde{e} \ g_0(\tilde{e},\tilde{t}), \quad
    E(\tilde{t}) = \int_0^{\infty} d\tilde{e} \ \tilde{e} g_0(\tilde{e}, \tilde{t}), \nonumber \\
    \langle \tilde{x}^2 \rangle(\tilde{t}) &= \int_0^\infty d\tilde{e} \ \frac{\tilde{e}}{2}\left(g_0(\tilde{e},\tilde{t}) + \frac{1}{2}g_1(\tilde{e},\tilde{t})\right), \nonumber \\
    \langle \tilde{k}^2 \rangle(\tilde{t}) &= \int_0^\infty d\tilde{e} \ \frac{\tilde{e}}{2}\left(g_0(\tilde{e},\tilde{t}) - \frac{1}{2}g_1(\tilde{e},\tilde{t})\right),
\label{eq:mode_expansion_observables}
\end{align}
where we also defined (twice) the total energy of the trapped gas $E(\tilde{t}) = \langle \tilde{x}^2+ \tilde{k}^2\rangle(\tilde{t})$.

\begin{figure}
    \centering
    \includegraphics[scale= 0.6]{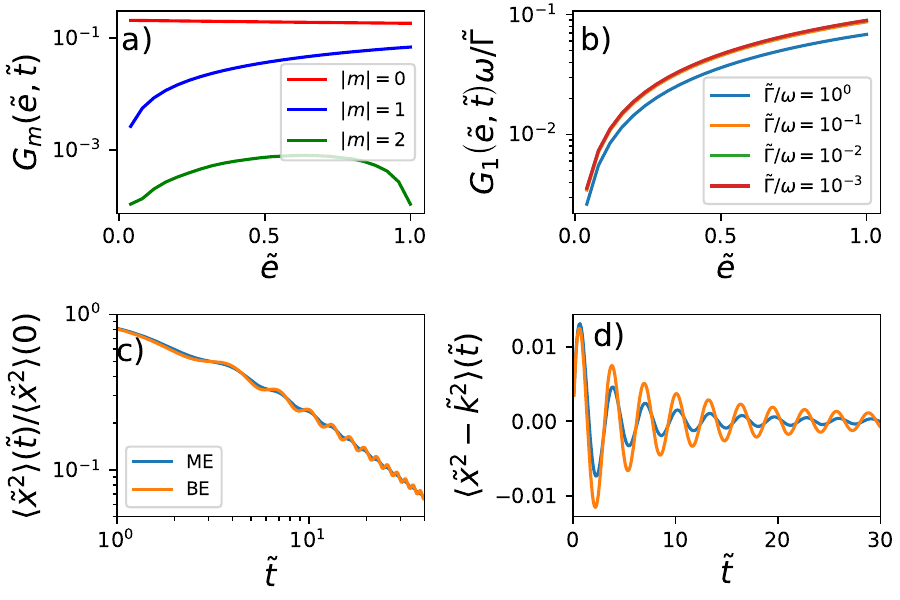}
    \caption{Amplitude of the $m$th mode $G_m(\tilde{e},\tilde{t})$ at $\tilde{t}= \tilde{\Gamma}/\omega$ for a) fixed  $\tilde{\Gamma}/\omega = 1$ and various $|m|$, and b) for fixed $|m|=1$ and various $\tilde{\Gamma}/\omega$. a) Modes with higher $m$ are suppressed at fixed $\tilde{\Gamma}/\omega$. b) In the limit of small $\tilde{\Gamma}/\omega$ and fixed $|m|$, $G_m(\tilde{e},\tilde{t})$ vanishes as $\tilde{\Gamma}/\omega$. This is evident from the data collapse upon rescaling $G_m(\tilde{e},\tilde{t})$. We compare the results for c) $\langle \tilde{x}^2\rangle$ and d) $\langle \tilde{x}^2 -\tilde{k}^2 \rangle$ calculated using the mode expansion (ME) and the Boltzmann equation (BE) in the presence of constant two-body losses with $\tilde{\Gamma}/\omega=1$.} 
    \label{fig:mode_expansion}
\end{figure}

The second benefit is that the mode expansion can be truncated at small values of $|m|$. This can be seen by noting that we initialise the system entirely in the $m=0$ mode. Populating higher modes is then due to the loss integral coupling the $m=0$ mode to a finite $m$. This coupling becomes readily suppressed at large $m$ as the $m$th mode becomes highly oscillatory, while the remainder of the integrand in Eq.~\eqref{eq:Am} is relatively smooth. The result is that the mode coupling diminishes with increasing $m$. To examine this issue more quantitatively we define the amplitude of the $|m|$th mode as:
\begin{equation}
    G_m(\tilde{e},\tilde{t}) = \left(g_m^2(\tilde{e},\tilde{t}) + g_{-m}^2(\tilde{e},\tilde{t})\right)^{1/2}.
    \label{eq:m_amp}
\end{equation}
In Fig.~(\ref{fig:mode_expansion}) we examinem $G_m(\tilde{e},\tilde{t})$ for various values $\tilde{\Gamma}/\omega$. One can immediately see that for fixed $\tilde{\Gamma}/\omega$, higher harmonics are indeed suppressed (a),  while each mode with $m\neq 0 $ is also suppressed as  $\tilde{\Gamma}/\omega \to 0$ (b). 
Hence the numerical analysis shows that the mode expansion can effectively be truncated at $|m|=1$ for small $\tilde{\Gamma}/\omega$ \footnote{We note that the mode expansion truncated at $m=0$ was reported in Ref.~\cite{Rosso22}}. In fact, it is evident that there is a sizeable $g_{\pm 1}(\tilde{e},\tilde{t})$ in Fig.~\eqref{fig:sketch}, whereas other periodicities in $\phi$ are indiscernible.

We further corroborate the validity of the mode expansion by 
comparing the dynamics of the potential energy $\langle \tilde{x}^2\rangle$ and the observable  $\langle\tilde{x}^2-\tilde{k}^2 \rangle$ calculated from the full Boltzmann equation, Eq.~(\ref{eq:Boltzmann}), and the mode expansion, Eq.~(\ref{eq:Boltzmann_mode_expansion}), truncated at $|m|=1$ in Fig.~(\ref{fig:mode_expansion}) (c) and (d). Both approaches qualitatively agree with one another; the mode expansion is accurate at capturing the overall decay of the potential energy and the frequency of the oscillations. However, there appears a discrepancy in the decay of the phase space quadrupole mode.
This discrepancy systematically decreases as the ratio $\tilde{\Gamma}/\omega$ is decreased, showing that the physics can be accurately captured by the truncated mode expansion. This is discussed further in Appendix \ref{app:mode_expansion}.

\section{Dynamics of a trapped Bose gas with Two-Body Losses}
\label{sec:dynamics}
\subsection{Dynamics of the Moments}
\begin{figure}
    \centering
    \includegraphics[scale = 0.55]{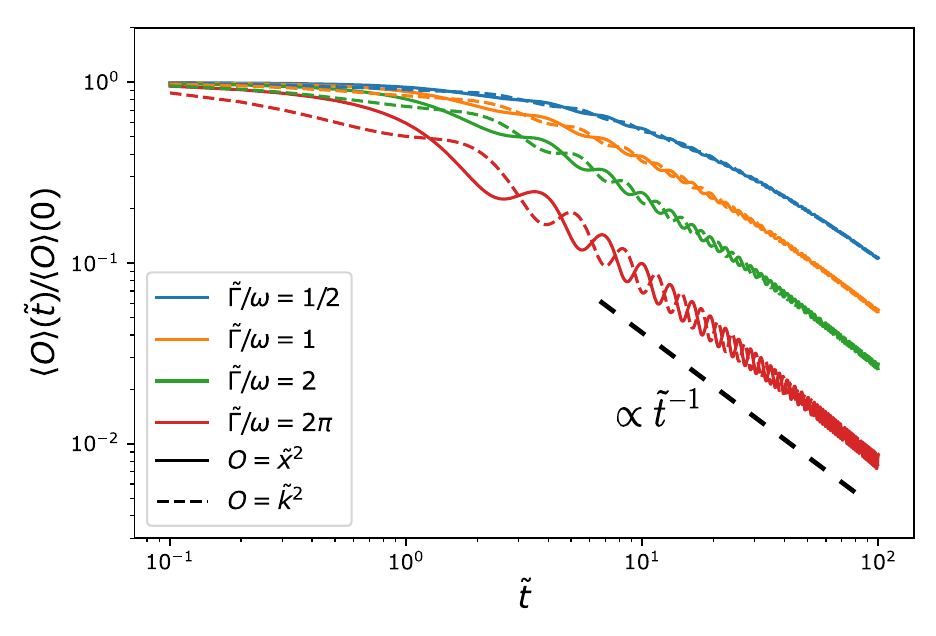}
    \caption{Results for the potential (kinetic) energy $O= \tilde{x}^2$ ($O= \tilde{k}^2$) from numerically evaluating Eq.~(\ref{eq:Boltzmann}) for various of $\tilde{\Gamma}/\omega$. Different colors indicate different values of the ratio $\tilde{\Gamma}/\omega$ while the potential and kinetic energies are denoted by the solid and dashed lines, respectively. The scaling behaviour of $\tilde{t}^{-1}$ is shown by the black dashed line. }
    \label{fig:xs_and_ks}
\end{figure}

First we consider the dynamics of a harmonically confined nearly hardcore Bose gas with constant two-body losses. In particular, we have examined the dynamics of the potential $\langle \tilde{x}^2 \rangle(t)$ and kinetic $\langle \tilde{k}^2 \rangle(t)$ energies in Fig.~(\ref{fig:xs_and_ks}). The dynamics of these observables exhibit two main features. The first is that there is an envelope that decays as $\tilde{t}^{-1}$ at long times. Given that these are extensive quantities, their decay is equivalent to the decrease in the overall number of particles \cite{Rosso22,Gerbino23}. The second feature is that there are persistent oscillations at exactly twice the trap frequency, $2\omega$. The amplitude of these oscillations decreases with decreasing $\tilde{\Gamma}/\omega$. In other words, if $\tilde{\Gamma}/\omega \to 0$, the system remains in equilibrium and there are no oscillations. In any case, these oscillations are the hallmark of a collective mode, and this shows that the gas begins to exhibit collective coherent dynamics after particle loss.

 Upon examining Eqs.~(\ref{eq:Boltzmann_mode_expansion}-\ref{eq:mode_expansion_observables}), one can immediately see that the the function $g_0(\tilde{e},\tilde{t})$ is responsible for the dynamics of the number of particles and the  overall decay of the potential and kinetic energies as $\tilde{t}^{-1}$. This was observed previously in Ref.~\cite{Rosso22}. The oscillations in the potential and kinetic energies are due to the dynamics of $g_1(\tilde{e},\tilde{t})$.  To isolate the oscillations, it is necessary to consider an observable that only depends on $g_1(\tilde{e},\tilde{t})$. We propose such an observable which we call the phase space quadrupole mode: 
\begin{equation}
    \langle \tilde{x}^2 - \tilde{k}^2 \rangle(\tilde{t}) = \int_0^\infty de \frac{e}{2} g_1(e,\tilde{t}).
    \label{eq:quad_mode}
\end{equation}
We note there is a second quadrupole mode, $\langle \tilde{x} \tilde{k} \rangle$, which is presented in Appendix \ref{app:mode_expansion}.
\begin{figure}
    \centering
    \includegraphics[scale = 0.58]{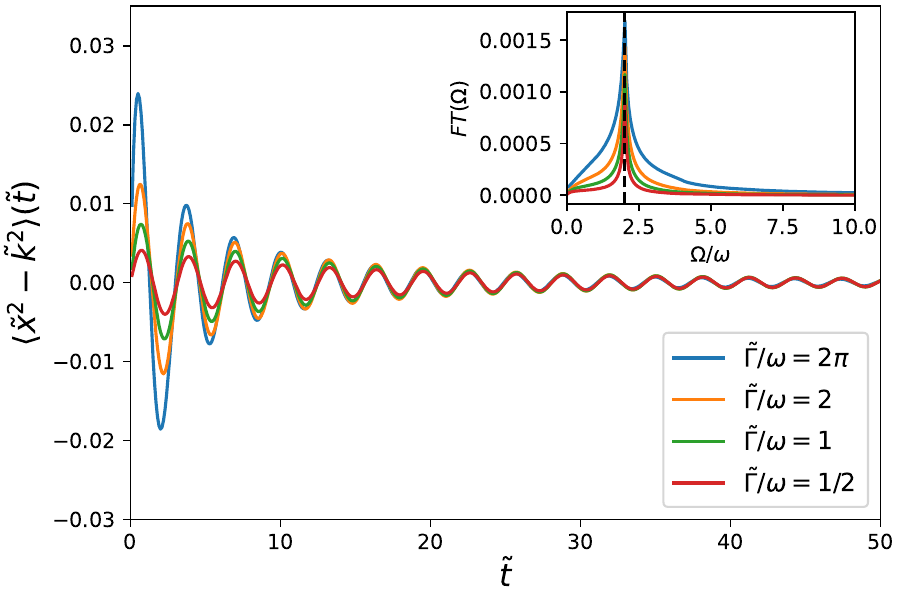}
    \caption{Phase space quadrupole mode, Eq.~(\ref{eq:quad_mode}) vs $\tilde{t}$ for various values of $\tilde{\Gamma}/\omega$. The phase space quadrupole mode only depends on $g_{1}(e,t)$ and as such does not contain the envelope $\propto \tilde{t}^{-1}$. The oscillations appear to be at a single frequency, $2\omega$.  The inset shows a fast Fourier transform of this data as a function of frequency $\Omega$. For each $\tilde{\Gamma}/\omega$ the Fourier transform reveals a single frequency $\Omega = 2\omega$, given by the black vertical dashed line.}
    \label{fig:quad_mode}
\end{figure}

The dynamics of this phase space quadrupole mode, as calculated from Eq.~(\ref{eq:Boltzmann}), is shown in Fig.~(\ref{fig:quad_mode}). This quantity oscillates at a single frequency $2\omega$ as evident from the structure of the Fourier transform in the inset. Although this frequency appears independent of $\tilde{\Gamma}/\omega$, we note this is due to the truncation of the expansion at the leading order of the dissipative dynamics \cite{Gerbino23,Rosso22} performed to obtain Eq.~\eqref{eq:Boltzmann}. We expect that the frequency of the oscillations will change as a function of the interaction $\Gamma$ if one includes finite elastic collisions between the fermionic modes scaling as $4g/(g^2+\gamma^2/4)$ \cite{Lenarcic18}. However, this effect is not present in the derivation of Eq.~\eqref{eq:Boltzmann} using a time-dependent generalized Gibbs ensemble \cite{Rossini21, Rosso22}.

\begin{figure}
    \centering
    \includegraphics[scale = 0.5]{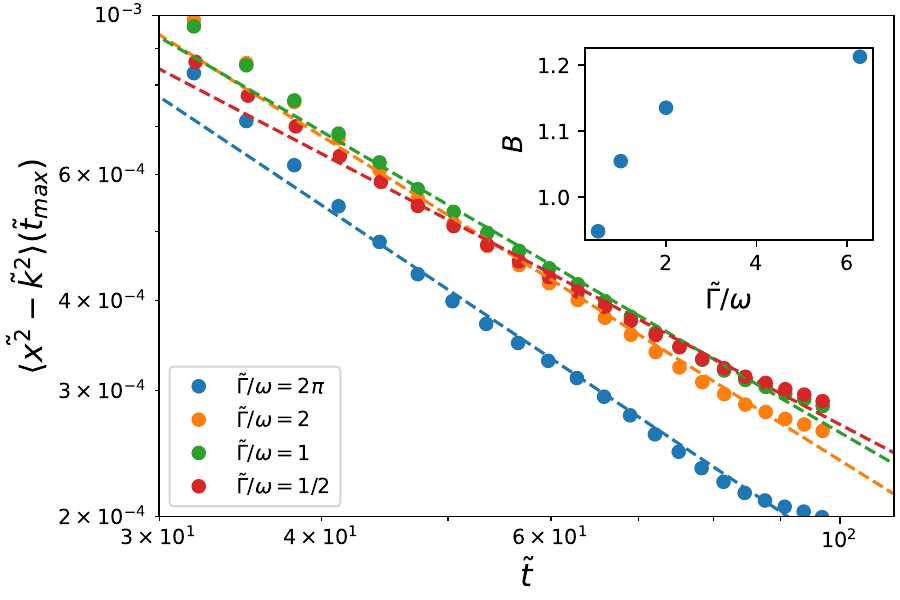}
    \caption{Decay of the maxima of the phase space quadrupole mode in the long-time limit for various two-body loss rates. Here we present the data on a log-log scale to highlight the power law decay. The dashed lines are power-law fits between $\tilde{t}=40-100$. The inset shows the power law exponent $B$ obtained from the fitting as a function of $\tilde{\Gamma}/\omega$.}
    \label{fig:amplitude_longtime}
\end{figure}

This long-time dynamics can be captured considering small  fluctuations $\delta g(\tilde{e},\tilde{\phi},\tilde{t})$ on top of a static distribution $\tilde{g}_0(e)$. 
The fluctuations have a spatial periodicity of $\cos(2\tilde{\phi})$ and oscillate a frequency $2\omega$ thus being a superposition of the $m=\pm1$ modes only. 
As shown in Appendix \ref{sec:long_time} the following ansatz:
\begin{equation}
g(\tilde{e},\tilde{\phi}, \tilde{t}) \approx N(\tilde{t}) \left(\tilde{g}_0(\tilde{e}) + \delta \tilde{g}(\tilde{e},\tilde{\phi}, \tilde{t})\right)
\label{eq:ansatz}
\end{equation}
is accurate at describing the long-time dynamics. Not only can it reproduce the overall $\tilde{t}^{-1}$ behaviour of the number of particles, but it states the collision integral vanishes as $\tilde{t}^{-1}$. Assuming that only the $m=\pm 1$ modes are dominant in the long time limit, one can derive an effective equation of motion for $G_1(\tilde{e},\tilde{t})$:

\begin{equation}
    \partial_t G_1(\tilde{e},\tilde{t}) \approx -\frac{B}{\tilde{t}}  G_1(\tilde{e},\tilde{t})
    \label{eq:toy_G}
\end{equation}
which yields $G_1(\tilde{e},\tilde{t}) \approx 1/\tilde{t}^{B}$, where $B$ is a constant that depends on $\Gamma/\omega$. In Fig.~\eqref{fig:amplitude_longtime} we focus on the long-time dynamics ($\tilde{t}\sim40-100$) of $\langle \tilde{x}^2-\tilde{k}^2\rangle(\tilde{t})$. As one can see, the decay of the phase-space quadrupole mode is well described by a power-law damping with an exponent $B\sim1$. The exponent as a function of $\Gamma/\omega$ is shown in the inset. Since the damping of the quadrupole mode follows a power-law, the damping rate in the long-time limit is predicted slows down as $1/\tilde{t}$. This is consistent with the observed slow-down of the decay of the quadrupole mode in Figs.~\eqref{fig:quad_mode}.

\subsection{Dynamics of the Density}
\begin{figure}
    \centering
    \includegraphics[scale = 0.58]{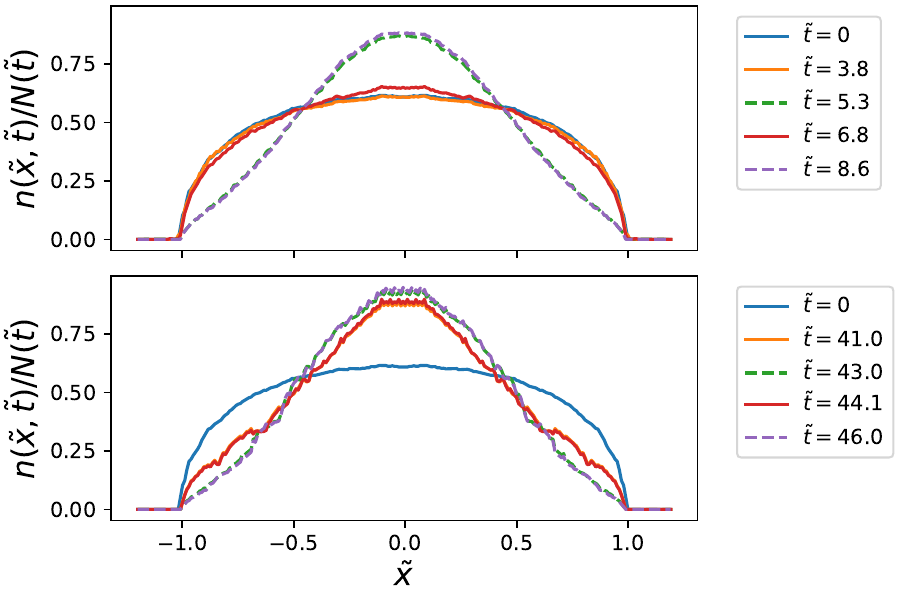}
    \caption{ Dynamics of the density for short times (top panel) and long times (bottom panel). The solid lines correspond to the maxima of the oscillations of $\langle \tilde{x}^2 \rangle(\tilde{t})$, while the dashed line correspond to the minima. At short times the maxima resemble the initial Thomas-Fermi distribution with small departures at the center, while the minima have a stronger concentration at the center of the trap. At long times, both the maximum and minimum distribution resemble one another. Similar plots exist for the rapidity distribution, but out of phase with the position space dynamics.}
    \label{fig:density_dynamics}
\end{figure}

Next we consider the dynamics of the density profile in real space. In Fig.~(\ref{fig:density_dynamics}) we present the results for the density at various times $\tilde{t}$ normalized to the instantaneous number of particles $N(\tilde{t})$. The top panel shows the density at short times, where each time corresponds to either a maximum (solid lines), or a minimum (dashed lines) of the oscillations in $\langle \tilde{x}^2\rangle(\tilde{t})$. Similarly, the bottom panel corresponds to later times. For short times, the density profile at a maximum of  $\langle \tilde{x}^2\rangle(\tilde{t})$ resembles the initial Thomas-Fermi distribution, while at a minimum it is more concentrated in the center and resembles a Gaussian.
At long times the amplitude of the quadrupole mode has decreased but it is still finite. Thus the difference between the density distribution at the maxima and the minima decreases up to small residual quadrupolar oscillations. These residual oscillations can also be described by Eq.~\eqref{eq:ansatz}, and are further evidence for the slow-down of the damping of the quadrupolar decay.

\subsection{Comparison to the Zero-Temperature Breathing Mode} Such dynamics in the density ought to be contrasted to the breathing mode for a hardcore Bose gas, i.e. in the absence of losses. In general, the breathing mode has been extensively studied in the context of 1D Bose gasses \cite{Menotti02,Moritz03,Minguzzi05,Fang14}. 
Primarily, we will be focused on a 1D Bose gas prepared initially in the equilibrium state at zero-temperature, Eq.~\eqref{eq:initial_state}, and then perform a quench in the trap. In this protocol, there are no losses, but the harmonic trap is quenched from $\omega_i$ to $\omega_f$. Under this condition it is well-known that the motion of the density profile in this case is self similar; the density profile at each moment is related to its initial value by a time-dependent rescaling of the position: 
\begin{equation}
    n(x,t) = \frac{1}{\lambda(t)} n\left(\frac{x}{\lambda(t)},0\right)
    \label{eq:density_dynamics_so21}
\end{equation}
where $\lambda^2(t) = \cos^2(\tilde{t}) + \frac{\omega_i^2}{\omega_f^2} \sin^2(\tilde{t})$. 

One can view these two protocols as the response of an initial state to a non-Hermitian perturbation (two-body losses) or a Hermitian one (quench of the trap frequencies). Although both protocols excite motion at a frequency $2\omega$, the density dynamics differ. This is evident from Fig.~\eqref{fig:density_dynamics} showing that in the lossy case the motion of the density profile is not self-similar in both the short and long-time limits.
The dynamics we observe is thus more reminiscent of the breathing at finite temperatures and finite interaction strengths, where self-similarity is also not observed \cite{Kheruntsyan2023}.

\section{Additional Protocols}
\label{sec:additional}
We further analyze the peculiar features of the LICM by considering two additional protocols. The first is where we consider constant two-body losses following the quench of the harmonic trap, and the second is the sudden removal of two body losses after a hold time $t_h$.

\subsection{Quenching the Harmonic Trap in the Presence of Losses}
We now investigate the dynamics of a strongly interacting Bose gas following a quench in the harmonic trapping potential from $\omega_i$ to $\omega_f$. The Boltzmann equation in dimensionless units is given by Eq.~(\ref{eq:Boltzmann}) with $\omega = \omega_f$. The main difference is the initial conditions. Assuming the system is initially in equilibrium at zero temperature in the initial harmonic trap, the initial distribution function has the form:
\begin{equation}
    f_{\tilde{k}}(\tilde{x},0) = \theta\left(1 -\tilde{k}^2 - \frac{\omega_i^2}{\omega_f^2} \tilde{x}^2\right).
\end{equation}
The initial distribution function is now anisotropic in phase space. As a result, not only the $m=0$ terms in the mode expansion is initially populated, but also terms with higher $m$.

\begin{figure}
    \centering
    \includegraphics[scale = 0.6]{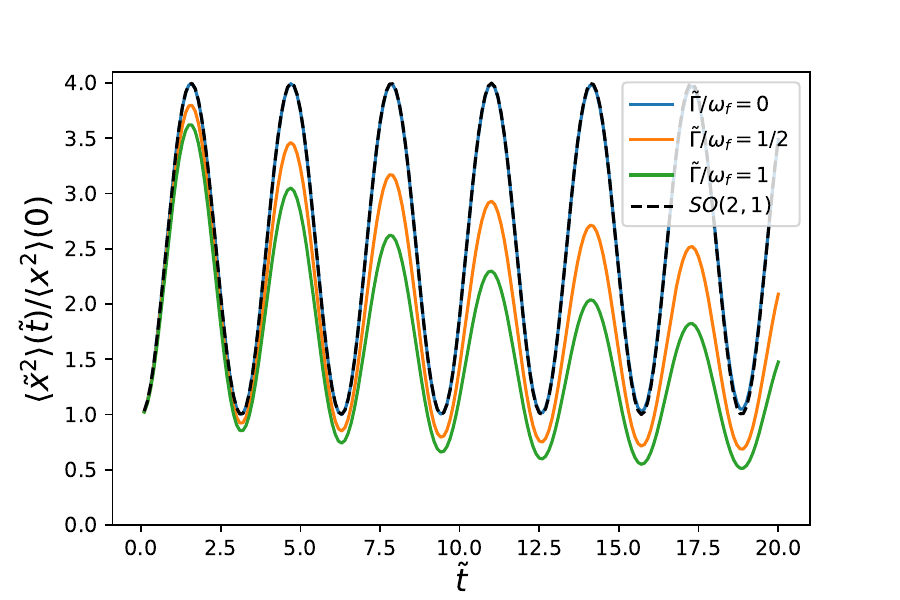}
    \caption{Dynamics of the moment of inertia, $\langle x^2\rangle(t)$ following a quench where $\omega_i/\omega_f =2$ including two-body losses. The dashed black line describes the predictions from SO(2,1) symmetry, i.e when $\tilde{\Gamma} = 0$.}
    \label{fig:quench_dynamics}
\end{figure}

In Fig.~(\ref{fig:quench_dynamics}) we present the results for a quench in the harmonic trap with $\omega_i/\omega_f = 2$. The oscillations in $\langle \tilde{x}^2 \rangle(t)$ still occur at a frequency of $2\omega_f$. The damping of these oscillations, unsurprisingly, depends on the strength of the dissipation, and becomes vanishingly small when $\Gamma/\omega \to 0$. 
When the two-body loss strength is zero, the system is governed by the unitary closed system dynamics described by the SO(2,1) symmetry. We have also confirmed in Fig.~(\ref{fig:quench_dynamics}) that our numerical solution in this limit matches the prediction from conformal symmetry:
\begin{align}
    \frac{\langle \tilde{x}^2 \rangle(\tilde{t})}{\langle \tilde{x}^2 \rangle(0)} &= \lambda^2(\tilde{t}), & \lambda^2(\tilde{t}) &= \cos^2(\tilde{t}) + \frac{\omega_i^2}{\omega_f^2} \sin^2(\tilde{t}).
    \label{eq:lambda(t)}
\end{align}

A key difference between the dynamics due to the quenched trap and the dynamics solely due to losses is the amplitude of the oscillations. In the absence of the quench, the amplitude of the oscillations is related to strength of the dissipation, as it is the dissipation that creates the non-equilibrium situation. In this case, after a quench, the system is initially in a highly excited non-equilibrium state. This generates oscillations with the amplitude set by the ratio $\omega_i/\omega_f$.

\subsection{The Sudden Removal of Losses}
\begin{figure}
    \centering
    \includegraphics[scale = 0.6]{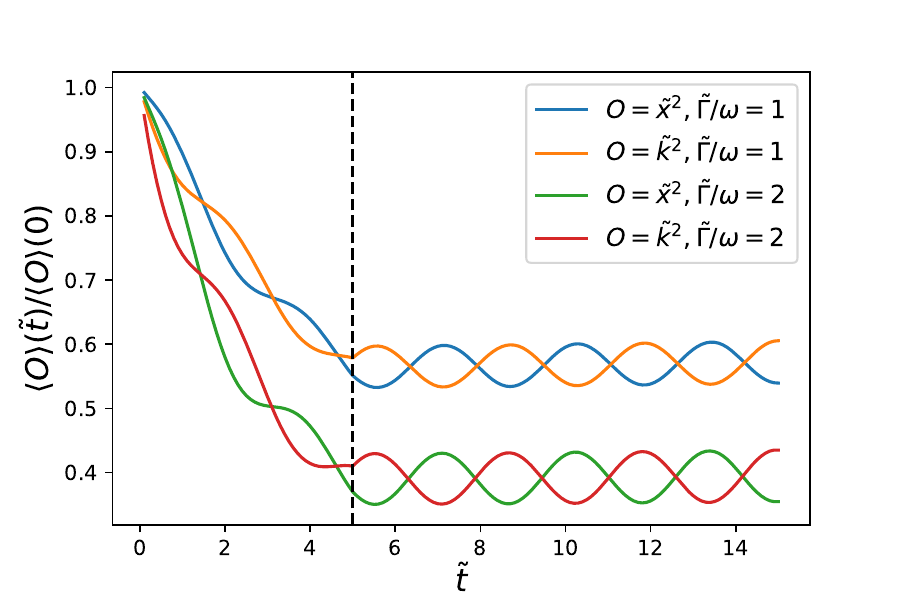}
    \caption{Dynamics of the potential and kinetic energies. The two-body losses are finite for $0<\tilde{t}<5$, while it is zero otherwise. For $\tilde{t}>\tilde{t}_h$ the system evolves under unitary dynamics which exhibit undamped oscillations at a frequency $2\omega$.}
    \label{fig:two_quench_moments}
\end{figure}
\begin{figure}
    \centering
    \includegraphics[scale=0.6]{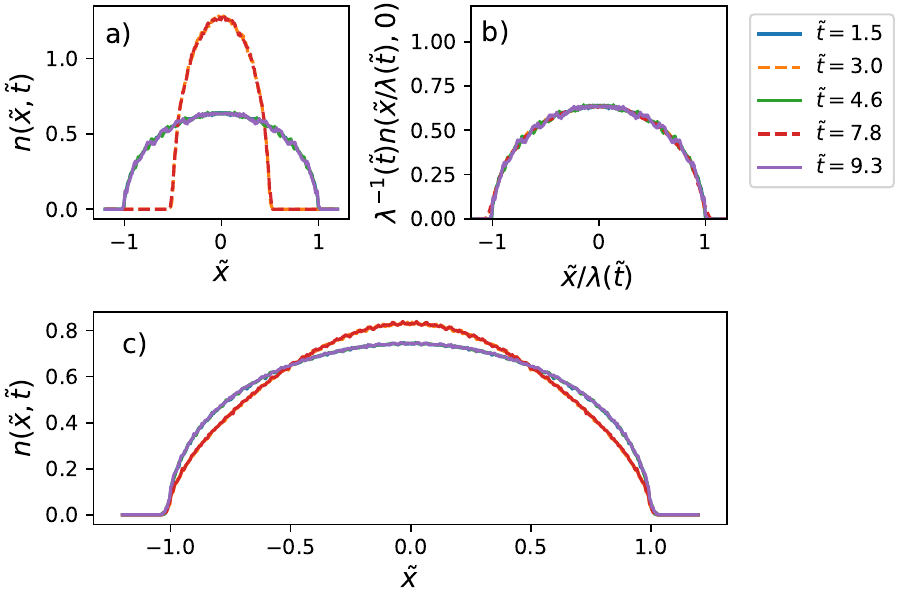}
    \caption{The dynamics of the density for a gas following a quench in the harmonic trapping potential in the absence of two-body losses (a) and b)) and following the removal of two-body losses after a hold time c). The solid (dashed) lines correspond to local maxima (minima) of the potential energy. Only in the absence of two-body losses are rescaling dynamics present.}
    \label{fig:density_comparsion_SM}
\end{figure}

In this protocol we suddenly turn on the two-body losses at $t=0$ and then we suddenly remove them at: $t=t_h$. For $0<t<t_h$, we expect the losses will lead to an overall decrease in the potential and kinetic energies, as well as oscillations as discussed in the previous section. For times $t>t_h$, the dynamics become unitary. In this case the dynamics will be governed by an emergent SO(2,1) symmetry \cite{Pitaevskii97,Minguzzi05, Werner06, Maki19, Maki20}. This SO(2,1) symmetry states that $\langle \tilde{x}^2\rangle$ will oscillate indefinitely at a frequency $2\omega$. 

We have performed numerical simulations for such a setup, and present the results in Fig.~(\ref{fig:two_quench_moments}) for the potential and kinetic energies with $\omega \, t_h = 5$. For $0<\tilde{t}<t_h$ one can clearly see the presence of losses and the overall $\tilde{t}^{-1}$ decay of the potential energy in the lossy regime, $0<t<t_h$, and then undamped periodic oscillations for $t>t_h$.

Although the dynamics of the potential energy and kinetic energy are fixed by the conformal symmetry, the dynamics of the density differs from the traditional breathing mode. This is shown in Fig.~(\ref{fig:density_comparsion_SM}). As discussed in the previous section, the dynamics of the density are self-similar, Eq.~\eqref{eq:density_dynamics_so21}, see Fig.~\eqref{fig:density_comparsion_SM} a-b). On the other hand, in the current experimental protocol, the dynamics of the density for $t>t_h$ are still not self-similar, Fig.~\eqref{fig:density_comparsion_SM} c).

Theoretically this difference can be described in terms of the initial conditions, i.e.\ the state of the gas when the unitary closed system dynamics begin. For the case of the quench in the harmonic trapping potential, the initial state (at $t=0$) is an equilibrium state within the initial harmonic trap. As discussed in Refs.~\cite{Maki19,Maki20}, such an initial state is diagonal in the basis of so-called conformal tower states. These conformal tower states are special as they have trivial rescaling dynamics. When the initial state is diagonal in said states, there are no interference effects between these states, and the dynamics of observables is completely governed by the trivial time-dependent rescaling dynamics. 

Contrast this to the case when we suddenly remove the losses. In this case the initial state (at $t_h$) is in a non-equilibrium state, and thus a state that cannot be expressed as a diagonal ensemble of conformal tower states. Hence there will be non-trivial interference effects in the dynamics of the density.

\section{Conclusions}
\label{sec:conc}
In this article we have examined the possibility of exciting collective modes in lossy 1D nearly hardcore Bose gasses. In particular, we considered a number of experimentally relevant protocols and found that in each case there can be oscillations in the potential energy at exactly twice the (final) trap frequency.
Such quantities are easy to access experimentally by comparing the dynamics of the width of the gas in position and and for the rapidities. 

The mechanism enabling such oscillation appears to be quite general. As shown in Appendix \ref{app:NI} equivalent physics occurs for a non-interacting Bose gas with two-body losses. We also expect this physics to occur for general $k$-body losses with $k>2$ \cite{Riggio23,Minganti2023dissipativephase, Perfetto23b}. Such losses would lead to different mode couplings, $\mathcal{A}_m$, in Eq.~(\ref{eq:Boltzmann_mode_expansion}), but the frequency of the oscillations will still be $2\omega$ to leading order in the loss strength. 

This work shows that the environment is a tool for probing quantum systems. As collective modes have played a fundamental role in characterising quantum gases, here we show that environment induced collective modes provide complementary and non-trivial information on these gasses. We leave for a future work a more detailed study of the specific properties of this environment-induced mode on the type of loss, finite elastic and inelastic collision strengths, emergent hydrodynamic behaviour and transport.

\section{Acknowledgements} 
We thank Stefano Giorgini and Sandro Stringari for useful discussions. 
We acknowledge financial support from Provincia Autonoma di Trento, European Union - NextGeneration EU, within PRIN 2022, PNRR M4C2, Project TANQU 2022FLSPAJ [CUP B53D23005130006].
AB would like to thank the Institut Henri Poincaré (UAR 839 CNRS-Sorbonne Université) and the LabEx CARMIN (ANR-10-LABX-59-01) for their support.

\appendix
\section{Deriving the Mode Expansion of the Boltzmann Equation}
\label{app:mode_expansion}
In this appendix we develop the mode expansion for the Boltzmann equation presented in the main text. Our starting point is the Lindblad master equation, Eq.~\eqref{eq:Lindblad}.
In the limit of strong elastic or inelastic interactions, the system can be described in terms of fermionic modes, or rapidities, which are weakly dissipative \cite{Zhu14, Syassen08,Duerr09, Rossini21,Rosso22, Rosso22b, Rosso23}. As shown in Ref.~\cite{Rosso22b} using the time-dependent generalized Gibbs ensemble and in Ref.~\cite{Perfetto23} using the Keldysh path integral that the dynamics of these system can be described by a Boltzmann equation for the distribution of these fermionic rapidities:
\begin{align}
    \left[\partial_{\tilde{t}}  + \left(\tilde{k}\partial_{\tilde{x}} - \tilde{x} \partial_{\tilde{k}}\right)\right]& f_{\tilde{k}}(\tilde{x},\tilde{t}) = \nonumber \\
    &- \frac{\Gamma}{\omega}\int_{\tilde{q},\tilde{k}} (\tilde{k}-\tilde{q})^2 f_{\tilde{k}}(\tilde{x},\tilde{t}) f_{\tilde{q}}(\tilde{x},\tilde{t}).
    \label{eq:boltzman_dimensionless}
\end{align}

Instead of working with $\tilde{x}$ and $\tilde{k}$ we work with modified polar coordinates:
\begin{align}
\tilde{e} &= \tilde{x}^2 + \tilde{k}^2, & \phi = \arctan\left( \frac{\tilde{k}}{\tilde{x}}\right).
\label{eq:polar_coordinates}
\end{align}
In these units the distribution transforms as: $f_{\tilde{k}}(\tilde{x},\tilde{t}) \rightarrow g(\tilde{e},\phi,\tilde{t}) = f_{\sqrt{\tilde{e}}\sin(\phi)}(\sqrt{\tilde{e}}\cos(\phi), \tilde{t})/2$. In terms of Eq.~(\ref{eq:polar_coordinates}) and $g(\tilde{e},\tilde{\phi},\tilde{t})$, the Jacobian for the transformation is unity, and we define: $\int_{\tilde{e}} =\int_0^{\infty} d\tilde{e}$ and $\int_{\tilde{\phi}} = \int_0^{2\pi} d\tilde{\phi}/(2\pi)$. The value of these units is that we can perform an expansion in terms of angular harmonics:
\begin{align}
    g(\tilde{e},\tilde{\phi},\tilde{t}) = \sum_{m=0}^{\infty} \left[g_m(\tilde{e},\tilde{t}) \cos(2m \tilde{\phi}) + g_{-m}(\tilde{e},\tilde{t}) \sin(2m\tilde{\phi})\right] \nonumber \\
    \label{eq:g}
\end{align}
where $m$ labels the harmonics. In principle, there are also harmonics with odd angular momentum. Harmonics with even (odd) angular momentum are even (odd) under a parity transformation, and since the Boltzmann equation also respects parity, these sectors are decoupled. Our analysis will be primarily concerned with the even harmonics.

The left hand side (LHS) of the Boltzmann equation in terms of Eq.~(\ref{eq:polar_coordinates}) can be straightforwardly evaluated by substituting Eq.~(\ref{eq:g}) into Eq.~(\ref{eq:boltzman_dimensionless}):
\begin{equation}
    {\rm LHS} = \partial_{\tilde{t}} g_{m}(\tilde{e},\tilde{t}) + 2m \omega g_{-m}(\tilde{e},\tilde{t}). 
    \label{eq:LHS}
\end{equation}
Note in order to obtain Eq.~(\ref{eq:LHS}) we note that $\cos(2m\tilde{\phi})$ and $\sin(2m\tilde{\phi})$ are orthogonal functions, and can be used to project out $g_m(\tilde{e},\tilde{t})$ and $g_{-m}(\tilde{e},\tilde{t})$, respectively. 

Although the expansion of the distribution function is clearest in terms of polar coordinates, it is more beneficial to work with modified coordinates $e$ and $\tilde{x}$ as the two-body loss term is local in $\tilde{x}$. In these coordinates: 

\begin{align}
    \cos(\tilde{\phi}) &= \frac{\tilde{x}}{\sqrt{\tilde{e}}}, & \sin(\tilde{\phi}) = \frac{\sqrt{\tilde{e}-\tilde{x}^2}}{\sqrt{\tilde{e}}}.
    \label{eq:modified_coordinates}
\end{align}
It is still possible to use the mode expansion in Eq.~(\ref{eq:g}), however, one needs to take care of the Jacobian for the transformation, and the definition of $\tilde{g}(\tilde{e},\tilde{x},\tilde{t})$ as the transformation from $\tilde{\phi} \to \tilde{x}$ is not single-valued. However, we can use the fact all the harmonics are even under parity to simplify the transformation. Thus we can define: 
\begin{align}
    \tilde{g}(\tilde{e},\tilde{x},\tilde{t}) &= g(\tilde{e},\tilde{\phi},\tilde{t}) + g(\tilde{e},\pi + \tilde{\phi}, \tilde{t}) &\nonumber \\
     \int_{\tilde{\phi}} & \rightarrow \int_{\tilde{x}}' = \int_{-\sqrt{\tilde{e}}}^{\sqrt{\tilde{e}}} d \tilde{x} \frac{1}{\pi \sqrt{\tilde{e}-\tilde{x}^2}},  &  \int_{\tilde{e}} &\rightarrow \int_{\tilde{e}}.
\end{align}

In these modified units, the LHS of the Boltzmann equation is the same, only the measure of the integrals changes. We now evaluate the RHS of Eq.~(\ref{eq:boltzman_dimensionless}) using the coordinates in Eq.~(\ref{eq:modified_coordinates}).  The result is:
\begin{align}
    {\rm RHS}&= -\frac{\tilde{\Gamma}}{\omega}\mathcal{A}_m, \nonumber \\
    \mathcal{A}_{m\geq0}&= (2-\delta_{|m|,0}) \int_0^1 d\tilde{e}' \int_{-min[\sqrt{\tilde{e}},\sqrt{\tilde{e}'}]}^{min[\sqrt{\tilde{e}},\sqrt{\tilde{e}'}]} dx \cos(2m\tilde{\phi})  \nonumber \\
    &\times\frac{1}{\pi^2}\left(\sqrt{\frac{\tilde{e}-\tilde{x}^2}{\tilde{e}'-\tilde{x}^2}} +\sqrt{\frac{\tilde{e}'-\tilde{x}^2}{\tilde{e}-\tilde{x}^2}} \right) \tilde{g}(\tilde{e},\tilde{x},\tilde{t})\tilde{g}(\tilde{e}',\tilde{x},\tilde{t}), \nonumber \\
    \mathcal{A}_{m<0}&= 2 \int_0^1 d\tilde{e}' \int_{-min[\sqrt{\tilde{e}},\sqrt{\tilde{e}'}]}^{min[\sqrt{\tilde{e}},\sqrt{\tilde{e}'}]} dx\sin(2|m|\tilde{\phi})\nonumber \\
    &\times\frac{1}{\pi^2}\left(\sqrt{\frac{\tilde{e}-\tilde{x}^2}{\tilde{e}'-\tilde{x}^2}} +\sqrt{\frac{\tilde{e}'-\tilde{x}^2}{\tilde{e}-\tilde{x}^2}} \right) \tilde{g}(\tilde{e},\tilde{x},\tilde{t})\tilde{g}(\tilde{e}',\tilde{x},\tilde{t}).
\end{align}
This follows from the orthogonality of the angular modes to project the results for a given $m$.

This mode expansion is in principle exact. As discussed in the main article, the utility of this approach is in truncating the expansion at some value of $|m|$. For example, the mode expansion at $m=0$ was presented in Ref.~\cite{Rosso22}. This mode expansion extends their result to higher harmonics.

\begin{figure}
    \centering
    \includegraphics[scale = 0.5]{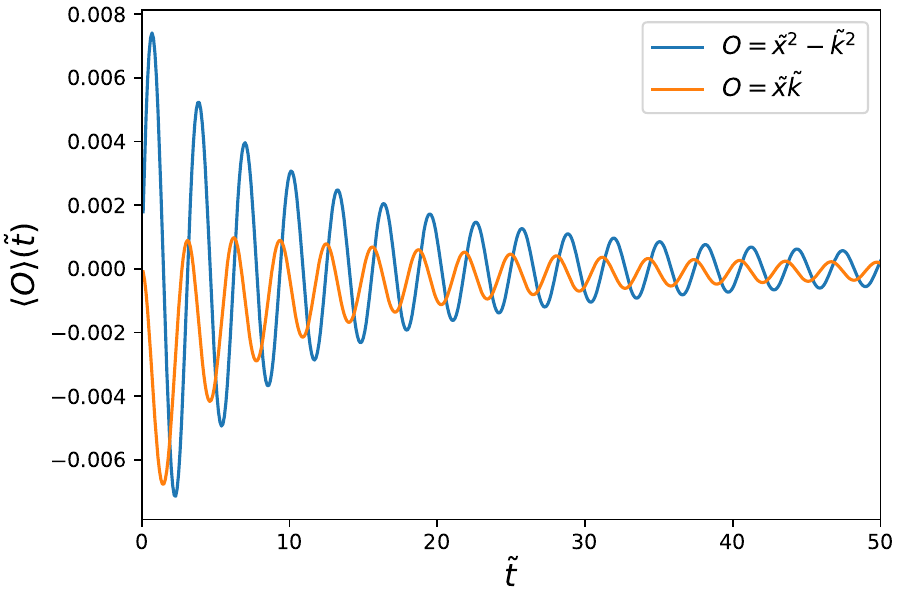}
    \caption{Comparison of the two quadrupole modes in Eqs.~(\ref{eq:quad_1}-\ref{eq:quad_2}) for $\tilde{\Gamma}/\omega=1$ calculated according to the Boltzmann equation, Eq.~\eqref{eq:boltzman_dimensionless}. }
    \label{fig:quad_modes_comp}
\end{figure}

The physical observables of interest in this can also be conveniently expressed in terms of the modes $m=0,\pm 1$:
\begin{align}
    N(\tilde{t}) &= \int_0^\infty d\tilde{e} \ g_0(\tilde{e},\tilde{t}), \nonumber \\
    E(\tilde{t}) &= \int_0^{\infty} d\tilde{e} \ \tilde{e} g_0(\tilde{e},\tilde{t}), \nonumber \\
    \langle \tilde{x}^2 \rangle(\tilde{t}) &= \int_0^\infty d\tilde{e} \frac{\tilde{e}}{2}\left(g_0(\tilde{e},\tilde{t}) + \frac{1}{2}g_1(\tilde{e},\tilde{t})\right), \nonumber \\
    \langle \tilde{k}^2 \rangle(\tilde{t}) &= \int_0^\infty d\tilde{e} \frac{\tilde{e}}{2}\left(g_0(\tilde{e},\tilde{t}) - \frac{1}{2}g_1(\tilde{e},\tilde{t})\right).
    \label{eq:mode_expansion_observables_app}
\end{align}
Eq.~\eqref{eq:mode_expansion_observables_app} also suggest there are two observables that only depend on $g_{\pm1}(\tilde{e},\tilde{t})$:
\begin{align}
    \langle \tilde{x}^2 - \tilde{k}^2\rangle(\tilde{t}) &= \int_{\tilde{e}} \frac{\tilde{e}}{2} g_1(\tilde{e},\tilde{t}), \label{eq:quad_1} \\
    \langle \tilde{x}\tilde{k}\rangle(\tilde{t}) &= \int_{\tilde{e}} \frac{\tilde{e}}{4}g_{-1}(\tilde{e},\tilde{t}). \label{eq:quad_2}
\end{align}
While Eq.~(\ref{eq:quad_1}) was discussed in the main text, here we present results for both Eq.~(\ref{eq:quad_1}-\ref{eq:quad_2}) for $\tilde{\Gamma}/\omega=1$. This is shown in Fig.~(\ref{fig:quad_modes_comp}). The dynamics of Eq.~(\ref{eq:quad_2}) are still periodic with a frequency of $2\omega$, but there is a $\pi/4$ phase shift with respect to Eq.~(\ref{eq:quad_1}). The asymmetry with respect to the zero-axis in Eq.~(\ref{eq:quad_2}) is due to the losses which are local in $x$.

\begin{figure}
    \centering
    \includegraphics[scale = 0.5]{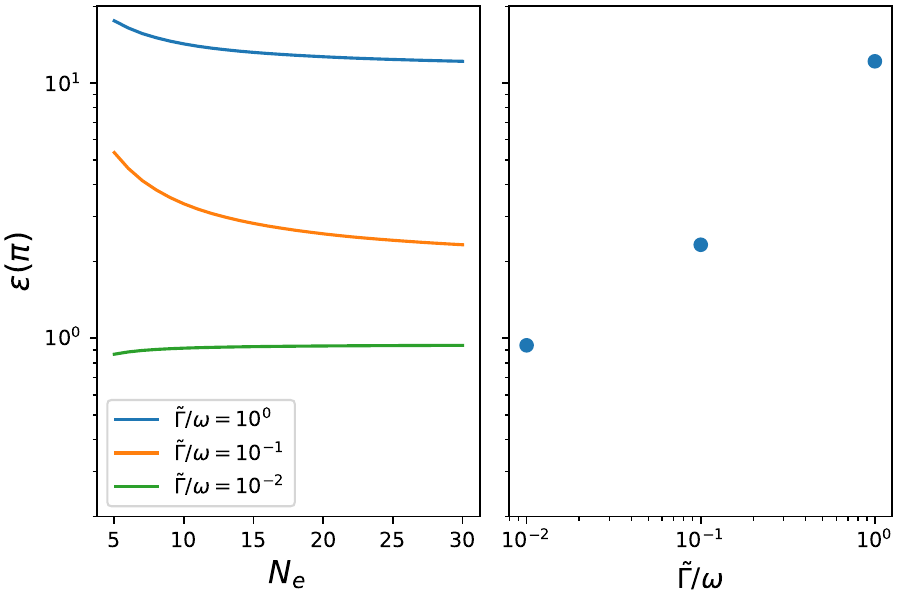}
    \caption{Comparison between full solution of the Boltzmann equation and the mode expansion. Here we plot the relative error, Eq.~\eqref{eq:rel_err} in the phase-space quadrupole mode evaluated at $\tilde{t} =\pi$. Here $N_e$ is the number of discretized energy steps used to evaluate the mode expansion. The inset shows the results for the relative error for $N_e=30$ and for various $\tilde{\Gamma}/\omega$. The relative error decreases as $\tilde{\Gamma}/\omega$ decreases.}
    \label{fig:me_be_comp}
\end{figure}

In Fig.~\eqref{fig:me_be_comp} we compare the results for the phase-space quadrupole mode evaluated at $\tilde{t} =\pi$ obtained by the truncated mode expansion with the results of the Boltzmann equation. When evaluating the mode expansion, we truncate the series at $m=\pm 1$ and discretize the energy into $N_e$ equidistant points. The relative error in this case is defined as: 
\begin{equation}
    \varepsilon(\tilde{t}) =\frac{\left| \langle \tilde{x}^2 - \tilde{k}^2\rangle_{\rm ME}(\tilde{t}) -  \langle \tilde{x}^2 - \tilde{k}^2\rangle_{\rm BE}(\tilde{t})\right|}{ \langle \tilde{x}^2 - \tilde{k}^2\rangle_{\rm BE}(\tilde{t})},
    \label{eq:rel_err}
\end{equation}
where ME and BE denote whether the phase-space quadrupole mode was calculated from the mode expansion or the full simulation of the Boltzmann equation, respectively. There are two-important trends. First, is that as we increase $N_e$ the relative error decreases as expected, and second, for fixed $N_e$, the relative error decreases with $\tilde{\Gamma}/\omega$. This provides evidence for the mode expansion being an effective way to model the dynamics.

\section{Analysis of Long-Time Limit}
\label{sec:long_time}
In the long-time limit our numerics suggest that the dynamics of the phase space distribution function can be written in a form:
\begin{equation}
g(\tilde{e}, \tilde{\phi}, \tilde{t}) \approx N(\tilde{t})\left(\tilde{g}_0(\tilde{e}) + \delta \tilde{g}(\tilde{e},\tilde{\phi},\tilde{t})\right).
\label{eqB:ansatz}
\end{equation}
In this appendix we show that this ansatz is consistent with 1) the overall decay in the number of atoms and 2) a slowly decaying quadrupole oscillation. To this end we substitute Eq.~\eqref{eqB:ansatz} into the Boltzmann equation and expand to linear order in $\tilde{\delta \tilde{g}}$. At zeroth order one obtains:
\begin{equation}
    \tilde{g}_0(\tilde{e}) \partial_{\tilde{t}} N(\tilde{t}) = -N^2(\tilde{t}) \mathcal{A}_0[\tilde{g}_0(\tilde{e})]
\end{equation}
where $\mathcal{A}_m[g]$ is given by Eq.~\eqref{eq:Am}. Dividing both sides by $\tilde{g}_0(\tilde{e},\tilde{t})$ and integrating the result over $\tilde{e}$ gives an equation of the form:
\begin{eqnarray}
    \partial_{\tilde{t}} N(\tilde{t}) = -C N^2(\tilde{t})
\end{eqnarray}
for some constant $C$. This naturally gives $N(\tilde{t}) \propto (C \tilde{t})^{-1}$.

We assume the linear order correction is due entirely to the quadrupole mode. The resulting equation for the linear order correction is:
\begin{align}
    \partial_t \delta \tilde{g}_m(\tilde{e},\tilde{\phi},\tilde{t}) &+2m\omega \delta \tilde{g}_{-m}(\tilde{e},\tilde{\phi},\tilde{t}) \nonumber \\
    &= -\frac{\partial_{\tilde{t}}N(\tilde{t})}{N({\tilde{t})}} \delta \tilde{g}_m(\tilde{e},\tilde{\phi},\tilde{t}) \nonumber\\ 
    &-N(\tilde{t}) \delta \mathcal{A}_m[\tilde{g}_0(\tilde{e})] \delta \tilde{g}_m(\tilde{e},\tilde{\phi},\tilde{t}) \nonumber \\
    &- N(\tilde{t}) \delta \mathcal{A}_m'[\delta\tilde{g}_m(\tilde{e},\tilde{\phi}, \tilde{t})] \delta \tilde{g}_0(\tilde{e})
    \label{eq:long_time_linear_order}
\end{align}
where $m = \pm 1 $ and we define the functions $\delta \mathcal{A}_m^{(')}$ which are related to the series expansion of $\mathcal{A}_m$ and depend solely on either $\tilde{g}_0(\tilde{e})$ or $\delta \tilde{g}_m(\tilde{e},\tilde{\phi},\tilde{t})$. The first line describes the oscillatory motion of the mode expansion. The damping can be qualitatively understood by noting that the right hand side  of Eq.~\eqref{eq:long_time_linear_order} decreases as $\tilde{t}^{-1}$. We expect the solution of Eq.~\eqref{eq:long_time_linear_order} to be oscillatory at a frequency $2\omega$ with a decaying envelope. The amplitude of the oscillations is captured by $G_1(\tilde{e},\tilde{t})$, defined in Eq.~\eqref{eq:m_amp}, will  approximately satisfy:
\begin{equation}
    \partial_t G_1(\tilde{e},\tilde{t}) \approx -\frac{B}{\tilde{t}} G_1(\tilde{e},\tilde{t})
    \label{eq:toy_G}
\end{equation}
which yields $G_1(\tilde{e},\tilde{t}) \approx 1/\tilde{t}^{B}$. Thus the damping rate in the long-time limit is predicted slows down as $1/\tilde{t}$.

We attempted to resolve this in our numerics by examining the decay of the maxima of the phase-space quadrupole mode. The results are presented in Fig.~\eqref{fig:amplitude_longtime}. It is clear that the maxima decay as an approximate power law with an exponent $B$ that is close to 1 (as shown in the inset). There also appear to be additional dynamics on top of the power-law decay of the maxima. Such terms should be described by Eq.~\eqref{eq:long_time_linear_order}. The power law decay, however, is well described by the simplified Eq.~\eqref{eq:toy_G}.

\begin{figure*}[h!t]
    \centering
    \includegraphics[scale = 0.9]{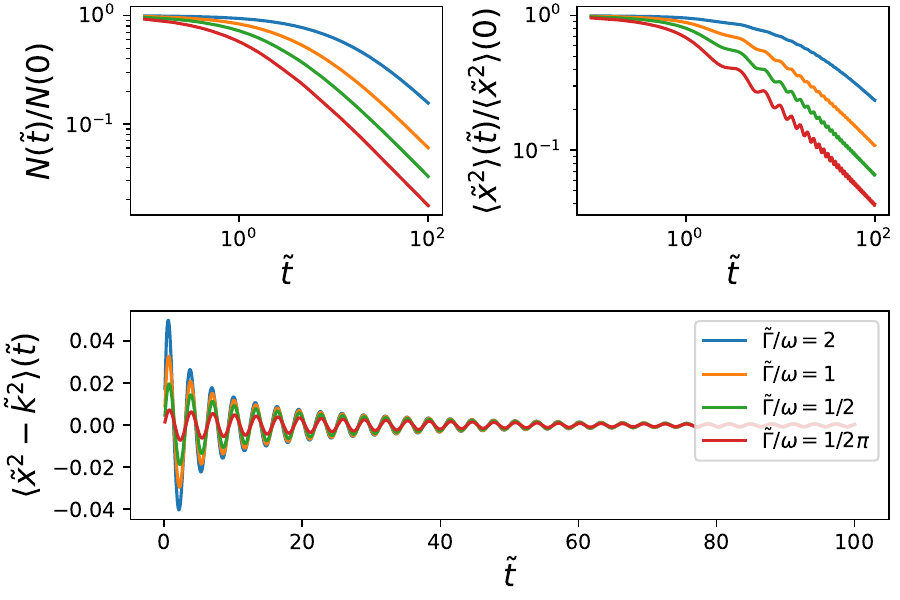}
    \caption{Dynamics of the total number (upper left), the moment of inertia (upper right), and phase space quadrupole mode (lower) for various $\tilde{\Gamma}/\omega$ for a non-interacting Bose gas with two-body losses. The behaviour of these observables is qualitatively similar to the Bose gas near the hardcore limit discussed in the main article. }
    \label{fig:non_int_fig}
\end{figure*}

\section{Two-Body Losses in the Non-Interacting Bose Gas}
\label{app:NI}

In this section we show that two-body losses can also induce oscillations in a non-interacting Bose gas. In order to derive the effective Boltzmann equation for weakly interacting bosons, we follow the methods of Ref.~\cite{Rosso22}. In this case we use a time-dependent ansatz for the density matrix:
\begin{equation}
    \rho(t) = \exp\left[-\sum_k \lambda_k(t) \psi_k^{\dagger}\psi_k\right]
\end{equation}
where $\psi_k^{(\dagger)}$ are the bosonic annihilation (creation operator) for the momentum mode $k$, with \mbox{$\psi(x) = \sum_k L^{-1/2} e^{i k x} \psi_k$}. We also define $\lambda_k(t)$ as a time-dependent Lagrange multiplier fixing the occupation of the $k$-th mode: $n_k(t) = \langle \psi_k^{\dagger}\psi_k\rangle(t)$. In fact, evaluating the equation of motion for the $n_k(t)$ using the Lindblad master equation, Eq.~\eqref{eq:Lindblad}, in the local density approximation gives the following Boltzmann equation:
\begin{equation}
    \left[\partial_t  + k \partial_x - \omega^2 x \partial_k \right]n_k(x,t) = -\Gamma \int_q n_q(x,t) n_k(x,t).
    \label{eq:boson_Boltzmann}
\end{equation}
In Eq.~(\ref{eq:boson_Boltzmann}) $n_k(x,t)$ is now the local density of bosons with momentum $k$. The LHS of Eq.~(\ref{eq:boson_Boltzmann}) again describes the evolution of the free bosons in a harmonic trap with frequency $\omega$, and is identical to the evolution of free  bosons (see LHS of Eq.~\eqref{eq:Boltzmann}). The RHS is the two-body losses term which differs from the RHS of Eq.~\eqref{eq:Boltzmann}. In contrast to the case of the fermionic rapidities, bosonic statistics do not forbid a contact interaction, thus we have a local two-body loss term without any momentum dependence. Again, we would like to note that this approach using a time-dependent ansatz for the density matrix coupled with the local density approximation provides identical results to the quantum kinetic equation derived perturbatively using Keldysh theory following the methods of Refs.~\cite{Gerbino23, Perfetto23,Huang23}.

In this analysis we consider all the bosons to be in their ground state:
\begin{equation}
    n_k(x,0) = 2N_0 e^{-x^2/a^2 - k^2 a^2}
\end{equation}
where $a$ is again the harmonic oscillator length, and $N_0$ is the initial particle number. We will work in units:
\begin{align}
    \tilde{x} &= \frac{x}{a}, & \tilde{k} &=ka, & \tilde{\Gamma} &= \frac{\Gamma}{a}, & \tilde{t} &= t \omega.
\end{align}
In terms of these units the Boltzmann equation becomes:
\begin{equation}
    \left[\partial_{\tilde{t}} + \tilde{k} \partial_{\tilde{x}} - \tilde{x} \partial_{\tilde{k}}\right]n_{\tilde{k}}(\tilde{x},\tilde{t}) = - \frac{\tilde{\Gamma}}{\omega}\int_{\tilde{q}} n_{\tilde{q}}(\tilde{x},\tilde{t})n_{\tilde{k}}(\tilde{x},\tilde{t}).
    \label{eq:boson_boltzmann_dimensionless}
\end{equation}

We will be focused on the following observables:

\begin{align}
    \frac{N(t)}{N_0} &= \int_{\tilde{x},\tilde{k}} n_{\tilde{k}}(\tilde{x}, \tilde{t}) & &\text{number} \nonumber \\
    \frac{\langle \tilde{x}^2\rangle(t)}{\langle \tilde{x}^2\rangle(0)} &= \int_{\tilde{x},\tilde{k}} \tilde{x}^2 n_{\tilde{k}}(\tilde{x}, \tilde{t}) & & \text{$2\times$ potential energy} \nonumber \\
    \frac{\langle \tilde{k}^2\rangle(t)}{\langle \tilde{k}^2\rangle(0)} &= \int_{\tilde{x},\tilde{k}} \tilde{k}^2 n_{\tilde{k}}(\tilde{x}, \tilde{t}) & & \text{$2\times$ kinetic energy}
\end{align}
The results for the numerical simulation of Eq.~(\ref{eq:boson_boltzmann_dimensionless}) are shown in Fig.~(\ref{fig:non_int_fig}) for the number, moment of inertia, and the phase space quadrupole mode. As in the case of a nearly hardcore Bose gas, there is an overall decay of the particle number, moment of inertia, and kinetic energy. On top of this decay, the moment of inertia and kinetic energy oscillate at a frequency of $2\omega$.

This can be understood from employing the mode expansion. The only difference between the mode expansion for non-interacting bosons and hardcore bosons is the form of the right hand side of the Boltzmann equation.

Finally we discuss the validity of our expansion about the non-interacting point. In this limit the elastic interaction strength and two-body loss rates are their bare values, $g$ and $\gamma$, respectively. In general we know that the effective strength of the interaction can be encoded in a single parameter: $\tilde{\gamma}(t) = |g-i\gamma|/(n(t))$, where we have made the density time-dependent. The non-interacting limit is when $\tilde{\gamma}(t) \ll1$. However, since the density is decreasing with a function of time, the system will inevitably flow to the strong coupling limit, where the system can be described in terms of hardcore bosons and the fermionic rapidities. Thus the kinetic equation in Eq.~\eqref{eq:boson_boltzmann_dimensionless} is only valid at intermediate times.

\bibliography{bibliography.bib} 

\begin{thebibliography}{31}
\expandafter\ifx\csname natexlab\endcsname\relax\def\natexlab#1{#1}\fi
\expandafter\ifx\csname bibnamefont\endcsname\relax
  \def\bibnamefont#1{#1}\fi
\expandafter\ifx\csname bibfnamefont\endcsname\relax
  \def\bibfnamefont#1{#1}\fi
\expandafter\ifx\csname citenamefont\endcsname\relax
  \def\citenamefont#1{#1}\fi
\expandafter\ifx\csname url\endcsname\relax
  \def\url#1{\texttt{#1}}\fi
\expandafter\ifx\csname urlprefix\endcsname\relax\def\urlprefix{URL }\fi
\providecommand{\bibinfo}[2]{#2}
\providecommand{\eprint}[2][]{\url{#2}}

\bibitem[{\citenamefont{Menotti and Stringari}(2002)}]{Menotti02}
\bibinfo{author}{\bibfnamefont{C.}~\bibnamefont{Menotti}} \bibnamefont{and} \bibinfo{author}{\bibfnamefont{S.}~\bibnamefont{Stringari}}, \bibinfo{journal}{Phys. Rev. A} \textbf{\bibinfo{volume}{66}}, \bibinfo{pages}{043610} (\bibinfo{year}{2002}), \urlprefix\url{https://link.aps.org/doi/10.1103/PhysRevA.66.043610}.

\bibitem[{\citenamefont{Minguzzi and Gangardt}(2005)}]{Minguzzi05}
\bibinfo{author}{\bibfnamefont{A.}~\bibnamefont{Minguzzi}} \bibnamefont{and} \bibinfo{author}{\bibfnamefont{D.~M.} \bibnamefont{Gangardt}}, \bibinfo{journal}{Phys. Rev. Lett.} \textbf{\bibinfo{volume}{94}}, \bibinfo{pages}{240404} (\bibinfo{year}{2005}), \urlprefix\url{https://link.aps.org/doi/10.1103/PhysRevLett.94.240404}.

\bibitem[{\citenamefont{Fang et~al.}(2014)\citenamefont{Fang, Carleo, Johnson, and Bouchoule}}]{Fang14}
\bibinfo{author}{\bibfnamefont{B.}~\bibnamefont{Fang}}, \bibinfo{author}{\bibfnamefont{G.}~\bibnamefont{Carleo}}, \bibinfo{author}{\bibfnamefont{A.}~\bibnamefont{Johnson}}, \bibnamefont{and} \bibinfo{author}{\bibfnamefont{I.}~\bibnamefont{Bouchoule}}, \bibinfo{journal}{Phys. Rev. Lett.} \textbf{\bibinfo{volume}{113}}, \bibinfo{pages}{035301} (\bibinfo{year}{2014}), \urlprefix\url{https://link.aps.org/doi/10.1103/PhysRevLett.113.035301}.

\bibitem[{\citenamefont{Pitaevskii and Stringari}(2016)}]{pitaevskii2016bose}
\bibinfo{author}{\bibfnamefont{L.}~\bibnamefont{Pitaevskii}} \bibnamefont{and} \bibinfo{author}{\bibfnamefont{S.}~\bibnamefont{Stringari}}, \emph{\bibinfo{title}{Bose-Einstein Condensation and Superfluidity}}, International Series of Monographs on Physics (\bibinfo{publisher}{OUP Oxford}, \bibinfo{year}{2016}), ISBN \bibinfo{isbn}{9780191076688}, \urlprefix\url{https://books.google.it/books?id=yHByCwAAQBAJa}.

\bibitem[{\citenamefont{Yamamoto et~al.}(2021)\citenamefont{Yamamoto, Nakagawa, Tsuji, Ueda, and Kawakami}}]{Yamamoto21}
\bibinfo{author}{\bibfnamefont{K.}~\bibnamefont{Yamamoto}}, \bibinfo{author}{\bibfnamefont{M.}~\bibnamefont{Nakagawa}}, \bibinfo{author}{\bibfnamefont{N.}~\bibnamefont{Tsuji}}, \bibinfo{author}{\bibfnamefont{M.}~\bibnamefont{Ueda}}, \bibnamefont{and} \bibinfo{author}{\bibfnamefont{N.}~\bibnamefont{Kawakami}}, \bibinfo{journal}{Phys. Rev. Lett.} \textbf{\bibinfo{volume}{127}}, \bibinfo{pages}{055301} (\bibinfo{year}{2021}), \urlprefix\url{https://link.aps.org/doi/10.1103/PhysRevLett.127.055301}.

\bibitem[{\citenamefont{Syassen et~al.}(2008)\citenamefont{Syassen, Bauer, Lettner, Volz, Dietze, García-Ripoll, Cirac, Rempe, and Dürr}}]{Syassen08}
\bibinfo{author}{\bibfnamefont{N.}~\bibnamefont{Syassen}}, \bibinfo{author}{\bibfnamefont{D.~M.} \bibnamefont{Bauer}}, \bibinfo{author}{\bibfnamefont{M.}~\bibnamefont{Lettner}}, \bibinfo{author}{\bibfnamefont{T.}~\bibnamefont{Volz}}, \bibinfo{author}{\bibfnamefont{D.}~\bibnamefont{Dietze}}, \bibinfo{author}{\bibfnamefont{J.~J.} \bibnamefont{García-Ripoll}}, \bibinfo{author}{\bibfnamefont{J.~I.} \bibnamefont{Cirac}}, \bibinfo{author}{\bibfnamefont{G.}~\bibnamefont{Rempe}}, \bibnamefont{and} \bibinfo{author}{\bibfnamefont{S.}~\bibnamefont{Dürr}}, \bibinfo{journal}{Science} \textbf{\bibinfo{volume}{320}}, \bibinfo{pages}{1329} (\bibinfo{year}{2008}).

\bibitem[{\citenamefont{D\"urr et~al.}(2009)\citenamefont{D\"urr, Garc\'{\i}a-Ripoll, Syassen, Bauer, Lettner, Cirac, and Rempe}}]{Duerr09}
\bibinfo{author}{\bibfnamefont{S.}~\bibnamefont{D\"urr}}, \bibinfo{author}{\bibfnamefont{J.~J.} \bibnamefont{Garc\'{\i}a-Ripoll}}, \bibinfo{author}{\bibfnamefont{N.}~\bibnamefont{Syassen}}, \bibinfo{author}{\bibfnamefont{D.~M.} \bibnamefont{Bauer}}, \bibinfo{author}{\bibfnamefont{M.}~\bibnamefont{Lettner}}, \bibinfo{author}{\bibfnamefont{J.~I.} \bibnamefont{Cirac}}, \bibnamefont{and} \bibinfo{author}{\bibfnamefont{G.}~\bibnamefont{Rempe}}, \bibinfo{journal}{Phys. Rev. A} \textbf{\bibinfo{volume}{79}}, \bibinfo{pages}{023614} (\bibinfo{year}{2009}).

\bibitem[{\citenamefont{Zhu et~al.}(2014)\citenamefont{Zhu, Gadway, Foss-Feig, Schachenmayer, Wall, Hazzard, Yan, Moses, Covey, Jin et~al.}}]{Zhu14}
\bibinfo{author}{\bibfnamefont{B.}~\bibnamefont{Zhu}}, \bibinfo{author}{\bibfnamefont{B.}~\bibnamefont{Gadway}}, \bibinfo{author}{\bibfnamefont{M.}~\bibnamefont{Foss-Feig}}, \bibinfo{author}{\bibfnamefont{J.}~\bibnamefont{Schachenmayer}}, \bibinfo{author}{\bibfnamefont{M.~L.} \bibnamefont{Wall}}, \bibinfo{author}{\bibfnamefont{K.~R.~A.} \bibnamefont{Hazzard}}, \bibinfo{author}{\bibfnamefont{B.}~\bibnamefont{Yan}}, \bibinfo{author}{\bibfnamefont{S.~A.} \bibnamefont{Moses}}, \bibinfo{author}{\bibfnamefont{J.~P.} \bibnamefont{Covey}}, \bibinfo{author}{\bibfnamefont{D.~S.} \bibnamefont{Jin}}, \bibnamefont{et~al.}, \bibinfo{journal}{Phys. Rev. Lett.} \textbf{\bibinfo{volume}{112}}, \bibinfo{pages}{070404} (\bibinfo{year}{2014}), \urlprefix\url{https://link.aps.org/doi/10.1103/PhysRevLett.112.070404}.

\bibitem[{\citenamefont{Johnson et~al.}(2017)\citenamefont{Johnson, Szigeti, Schemmer, and Bouchoule}}]{Johnson_2017}
\bibinfo{author}{\bibfnamefont{A.}~\bibnamefont{Johnson}}, \bibinfo{author}{\bibfnamefont{S.~S.} \bibnamefont{Szigeti}}, \bibinfo{author}{\bibfnamefont{M.}~\bibnamefont{Schemmer}}, \bibnamefont{and} \bibinfo{author}{\bibfnamefont{I.}~\bibnamefont{Bouchoule}}, \bibinfo{journal}{Phys. Rev. A} \textbf{\bibinfo{volume}{96}}, \bibinfo{pages}{013623} (\bibinfo{year}{2017}), \urlprefix\url{https://link.aps.org/doi/10.1103/PhysRevA.96.013623}.

\bibitem[{\citenamefont{Tomita et~al.}(2017)\citenamefont{Tomita, Nakajima, Danshita, Takasu, and Takahashi}}]{Tomita17}
\bibinfo{author}{\bibfnamefont{T.}~\bibnamefont{Tomita}}, \bibinfo{author}{\bibfnamefont{S.}~\bibnamefont{Nakajima}}, \bibinfo{author}{\bibfnamefont{I.}~\bibnamefont{Danshita}}, \bibinfo{author}{\bibfnamefont{Y.}~\bibnamefont{Takasu}}, \bibnamefont{and} \bibinfo{author}{\bibfnamefont{Y.}~\bibnamefont{Takahashi}}, \bibinfo{journal}{Science Advances} \textbf{\bibinfo{volume}{3}}, \bibinfo{pages}{e1701513} (\bibinfo{year}{2017}), \eprint{https://www.science.org/doi/pdf/10.1126/sciadv.1701513}, \urlprefix\url{https://www.science.org/doi/abs/10.1126/sciadv.1701513}.

\bibitem[{\citenamefont{Bouchoule and Schemmer}(2020)}]{Bouchoule20}
\bibinfo{author}{\bibfnamefont{I.}~\bibnamefont{Bouchoule}} \bibnamefont{and} \bibinfo{author}{\bibfnamefont{M.}~\bibnamefont{Schemmer}}, \bibinfo{journal}{SciPost Phys.} \textbf{\bibinfo{volume}{8}}, \bibinfo{pages}{060} (\bibinfo{year}{2020}), \urlprefix\url{https://scipost.org/10.21468/SciPostPhys.8.4.060}.

\bibitem[{\citenamefont{Bouchoule et~al.}(2020)\citenamefont{Bouchoule, Doyon, and Dubail}}]{Bouchoule20b}
\bibinfo{author}{\bibfnamefont{I.}~\bibnamefont{Bouchoule}}, \bibinfo{author}{\bibfnamefont{B.}~\bibnamefont{Doyon}}, \bibnamefont{and} \bibinfo{author}{\bibfnamefont{J.}~\bibnamefont{Dubail}}, \bibinfo{journal}{SciPost Phys.} \textbf{\bibinfo{volume}{9}}, \bibinfo{pages}{044} (\bibinfo{year}{2020}), \urlprefix\url{https://scipost.org/10.21468/SciPostPhys.9.4.044}.

\bibitem[{\citenamefont{Bouchoule and Dubail}(2021)}]{Bouchoule21}
\bibinfo{author}{\bibfnamefont{I.}~\bibnamefont{Bouchoule}} \bibnamefont{and} \bibinfo{author}{\bibfnamefont{J.}~\bibnamefont{Dubail}}, \bibinfo{journal}{Phys. Rev. Lett.} \textbf{\bibinfo{volume}{126}}, \bibinfo{pages}{160603} (\bibinfo{year}{2021}), \urlprefix\url{https://link.aps.org/doi/10.1103/PhysRevLett.126.160603}.

\bibitem[{\citenamefont{Rossini et~al.}(2021)\citenamefont{Rossini, Ghermaoui, Aguilera, Vatr\'e, Bouganne, Beugnon, Gerbier, and Mazza}}]{Rossini21}
\bibinfo{author}{\bibfnamefont{D.}~\bibnamefont{Rossini}}, \bibinfo{author}{\bibfnamefont{A.}~\bibnamefont{Ghermaoui}}, \bibinfo{author}{\bibfnamefont{M.~B.} \bibnamefont{Aguilera}}, \bibinfo{author}{\bibfnamefont{R.}~\bibnamefont{Vatr\'e}}, \bibinfo{author}{\bibfnamefont{R.}~\bibnamefont{Bouganne}}, \bibinfo{author}{\bibfnamefont{J.}~\bibnamefont{Beugnon}}, \bibinfo{author}{\bibfnamefont{F.}~\bibnamefont{Gerbier}}, \bibnamefont{and} \bibinfo{author}{\bibfnamefont{L.}~\bibnamefont{Mazza}}, \bibinfo{journal}{Phys. Rev. A} \textbf{\bibinfo{volume}{103}}, \bibinfo{pages}{L060201} (\bibinfo{year}{2021}), \urlprefix\url{https://link.aps.org/doi/10.1103/PhysRevA.103.L060201}.

\bibitem[{\citenamefont{Rosso et~al.}(2022{\natexlab{a}})\citenamefont{Rosso, Biella, and Mazza}}]{Rosso22}
\bibinfo{author}{\bibfnamefont{L.}~\bibnamefont{Rosso}}, \bibinfo{author}{\bibfnamefont{A.}~\bibnamefont{Biella}}, \bibnamefont{and} \bibinfo{author}{\bibfnamefont{L.}~\bibnamefont{Mazza}}, \bibinfo{journal}{SciPost Phys.} \textbf{\bibinfo{volume}{12}}, \bibinfo{pages}{044} (\bibinfo{year}{2022}{\natexlab{a}}), \urlprefix\url{https://scipost.org/10.21468/SciPostPhys.12.1.044}.

\bibitem[{\citenamefont{Rosso et~al.}(2022{\natexlab{b}})\citenamefont{Rosso, Mazza, and Biella}}]{Rosso22b}
\bibinfo{author}{\bibfnamefont{L.}~\bibnamefont{Rosso}}, \bibinfo{author}{\bibfnamefont{L.}~\bibnamefont{Mazza}}, \bibnamefont{and} \bibinfo{author}{\bibfnamefont{A.}~\bibnamefont{Biella}}, \bibinfo{journal}{Phys. Rev. A} \textbf{\bibinfo{volume}{105}}, \bibinfo{pages}{L051302} (\bibinfo{year}{2022}{\natexlab{b}}), \urlprefix\url{https://link.aps.org/doi/10.1103/PhysRevA.105.L051302}.

\bibitem[{\citenamefont{Rosso et~al.}(2023)\citenamefont{Rosso, Biella, De~Nardis, and Mazza}}]{Rosso23}
\bibinfo{author}{\bibfnamefont{L.}~\bibnamefont{Rosso}}, \bibinfo{author}{\bibfnamefont{A.}~\bibnamefont{Biella}}, \bibinfo{author}{\bibfnamefont{J.}~\bibnamefont{De~Nardis}}, \bibnamefont{and} \bibinfo{author}{\bibfnamefont{L.}~\bibnamefont{Mazza}}, \bibinfo{journal}{Phys. Rev. A} \textbf{\bibinfo{volume}{107}}, \bibinfo{pages}{013303} (\bibinfo{year}{2023}), \urlprefix\url{https://link.aps.org/doi/10.1103/PhysRevA.107.013303}.

\bibitem[{\citenamefont{Liu et~al.}(2022)\citenamefont{Liu, Shi, and Wang}}]{liu2022weakly}
\bibinfo{author}{\bibfnamefont{C.}~\bibnamefont{Liu}}, \bibinfo{author}{\bibfnamefont{Z.-Y.} \bibnamefont{Shi}}, \bibnamefont{and} \bibinfo{author}{\bibfnamefont{C.}~\bibnamefont{Wang}}, \emph{\bibinfo{title}{Weakly interacting bose gas with two-body losses}} (\bibinfo{year}{2022}), \eprint{2209.10427}.

\bibitem[{\citenamefont{Gerbino et~al.}(2023)\citenamefont{Gerbino, Lesanovsky, and Perfetto}}]{Gerbino23}
\bibinfo{author}{\bibfnamefont{F.}~\bibnamefont{Gerbino}}, \bibinfo{author}{\bibfnamefont{I.}~\bibnamefont{Lesanovsky}}, \bibnamefont{and} \bibinfo{author}{\bibfnamefont{G.}~\bibnamefont{Perfetto}}, \emph{\bibinfo{title}{Large-scale universality in quantum reaction-diffusion from keldysh field theory}} (\bibinfo{year}{2023}), \eprint{arXiv:2307.14945}.

\bibitem[{\citenamefont{Perfetto et~al.}(2023{\natexlab{a}})\citenamefont{Perfetto, Carollo, Garrahan, and Lesanovsky}}]{Perfetto23}
\bibinfo{author}{\bibfnamefont{G.}~\bibnamefont{Perfetto}}, \bibinfo{author}{\bibfnamefont{F.}~\bibnamefont{Carollo}}, \bibinfo{author}{\bibfnamefont{J.~P.} \bibnamefont{Garrahan}}, \bibnamefont{and} \bibinfo{author}{\bibfnamefont{I.}~\bibnamefont{Lesanovsky}}, \bibinfo{journal}{Phys. Rev. Lett.} \textbf{\bibinfo{volume}{130}}, \bibinfo{pages}{210402} (\bibinfo{year}{2023}{\natexlab{a}}), \urlprefix\url{https://link.aps.org/doi/10.1103/PhysRevLett.130.210402}.

\bibitem[{\citenamefont{Riggio et~al.}(2024)\citenamefont{Riggio, Rosso, Karevski, and Dubail}}]{Riggio23}
\bibinfo{author}{\bibfnamefont{F.}~\bibnamefont{Riggio}}, \bibinfo{author}{\bibfnamefont{L.}~\bibnamefont{Rosso}}, \bibinfo{author}{\bibfnamefont{D.}~\bibnamefont{Karevski}}, \bibnamefont{and} \bibinfo{author}{\bibfnamefont{J.}~\bibnamefont{Dubail}}, \bibinfo{journal}{Phys. Rev. A} \textbf{\bibinfo{volume}{109}}, \bibinfo{pages}{023311} (\bibinfo{year}{2024}), \urlprefix\url{https://link.aps.org/doi/10.1103/PhysRevA.109.023311}.

\bibitem[{\citenamefont{Huang et~al.}(2023)\citenamefont{Huang, Giamarchi, and Cazalilla}}]{Huang23}
\bibinfo{author}{\bibfnamefont{C.-H.} \bibnamefont{Huang}}, \bibinfo{author}{\bibfnamefont{T.}~\bibnamefont{Giamarchi}}, \bibnamefont{and} \bibinfo{author}{\bibfnamefont{M.~A.} \bibnamefont{Cazalilla}}, \bibinfo{journal}{Phys. Rev. Res.} \textbf{\bibinfo{volume}{5}}, \bibinfo{pages}{043192} (\bibinfo{year}{2023}), \urlprefix\url{https://link.aps.org/doi/10.1103/PhysRevResearch.5.043192}.

\bibitem[{\citenamefont{Lenar\ifmmode \check{c}\else \v{c}\fi{}i\ifmmode~\check{c}\else \v{c}\fi{} et~al.}(2018)\citenamefont{Lenar\ifmmode \check{c}\else \v{c}\fi{}i\ifmmode~\check{c}\else \v{c}\fi{}, Lange, and Rosch}}]{Lenarcic18}
\bibinfo{author}{\bibfnamefont{Z.}~\bibnamefont{Lenar\ifmmode \check{c}\else \v{c}\fi{}i\ifmmode~\check{c}\else \v{c}\fi{}}}, \bibinfo{author}{\bibfnamefont{F.}~\bibnamefont{Lange}}, \bibnamefont{and} \bibinfo{author}{\bibfnamefont{A.}~\bibnamefont{Rosch}}, \bibinfo{journal}{Phys. Rev. B} \textbf{\bibinfo{volume}{97}}, \bibinfo{pages}{024302} (\bibinfo{year}{2018}), \urlprefix\url{https://link.aps.org/doi/10.1103/PhysRevB.97.024302}.

\bibitem[{\citenamefont{Moritz et~al.}(2003)\citenamefont{Moritz, St\"oferle, K\"ohl, and Esslinger}}]{Moritz03}
\bibinfo{author}{\bibfnamefont{H.}~\bibnamefont{Moritz}}, \bibinfo{author}{\bibfnamefont{T.}~\bibnamefont{St\"oferle}}, \bibinfo{author}{\bibfnamefont{M.}~\bibnamefont{K\"ohl}}, \bibnamefont{and} \bibinfo{author}{\bibfnamefont{T.}~\bibnamefont{Esslinger}}, \bibinfo{journal}{Phys. Rev. Lett.} \textbf{\bibinfo{volume}{91}}, \bibinfo{pages}{250402} (\bibinfo{year}{2003}), \urlprefix\url{https://link.aps.org/doi/10.1103/PhysRevLett.91.250402}.

\bibitem[{\citenamefont{Francis A.~Bayocboc and Kheruntsyan}(2023)}]{Kheruntsyan2023}
\bibinfo{author}{\bibfnamefont{J.}~\bibnamefont{Francis A.~Bayocboc}} \bibnamefont{and} \bibinfo{author}{\bibfnamefont{K.~V.} \bibnamefont{Kheruntsyan}}, \bibinfo{journal}{Comptes Rendus. Physique}  (\bibinfo{year}{2023}), \bibinfo{note}{online first}.

\bibitem[{\citenamefont{Pitaevskii and Rosch}(1997)}]{Pitaevskii97}
\bibinfo{author}{\bibfnamefont{L.~P.} \bibnamefont{Pitaevskii}} \bibnamefont{and} \bibinfo{author}{\bibfnamefont{A.}~\bibnamefont{Rosch}}, \bibinfo{journal}{Phys. Rev. A} \textbf{\bibinfo{volume}{55}}, \bibinfo{pages}{R853} (\bibinfo{year}{1997}), \urlprefix\url{https://link.aps.org/doi/10.1103/PhysRevA.55.R853}.

\bibitem[{\citenamefont{Werner and Castin}(2006)}]{Werner06}
\bibinfo{author}{\bibfnamefont{F.}~\bibnamefont{Werner}} \bibnamefont{and} \bibinfo{author}{\bibfnamefont{Y.}~\bibnamefont{Castin}}, \bibinfo{journal}{Phys. Rev. A} \textbf{\bibinfo{volume}{74}}, \bibinfo{pages}{053604} (\bibinfo{year}{2006}), \urlprefix\url{https://link.aps.org/doi/10.1103/PhysRevA.74.053604}.

\bibitem[{\citenamefont{Maki and Zhou}(2019)}]{Maki19}
\bibinfo{author}{\bibfnamefont{J.}~\bibnamefont{Maki}} \bibnamefont{and} \bibinfo{author}{\bibfnamefont{F.}~\bibnamefont{Zhou}}, \bibinfo{journal}{Phys. Rev. A} \textbf{\bibinfo{volume}{100}}, \bibinfo{pages}{023601} (\bibinfo{year}{2019}), \urlprefix\url{https://link.aps.org/doi/10.1103/PhysRevA.100.023601}.

\bibitem[{\citenamefont{Maki and Zhou}(2020)}]{Maki20}
\bibinfo{author}{\bibfnamefont{J.}~\bibnamefont{Maki}} \bibnamefont{and} \bibinfo{author}{\bibfnamefont{F.}~\bibnamefont{Zhou}}, \bibinfo{journal}{Phys. Rev. A} \textbf{\bibinfo{volume}{102}}, \bibinfo{pages}{063319} (\bibinfo{year}{2020}), \urlprefix\url{https://link.aps.org/doi/10.1103/PhysRevA.102.063319}.

\bibitem[{\citenamefont{Minganti et~al.}(2023)\citenamefont{Minganti, Savona, and Biella}}]{Minganti2023dissipativephase}
\bibinfo{author}{\bibfnamefont{F.}~\bibnamefont{Minganti}}, \bibinfo{author}{\bibfnamefont{V.}~\bibnamefont{Savona}}, \bibnamefont{and} \bibinfo{author}{\bibfnamefont{A.}~\bibnamefont{Biella}}, \bibinfo{journal}{{Quantum}} \textbf{\bibinfo{volume}{7}}, \bibinfo{pages}{1170} (\bibinfo{year}{2023}), ISSN \bibinfo{issn}{2521-327X}, \urlprefix\url{https://doi.org/10.22331/q-2023-11-07-1170}.

\bibitem[{\citenamefont{Perfetto et~al.}(2023{\natexlab{b}})\citenamefont{Perfetto, Carollo, Garrahan, and Lesanovsky}}]{Perfetto23b}
\bibinfo{author}{\bibfnamefont{G.}~\bibnamefont{Perfetto}}, \bibinfo{author}{\bibfnamefont{F.}~\bibnamefont{Carollo}}, \bibinfo{author}{\bibfnamefont{J.~P.} \bibnamefont{Garrahan}}, \bibnamefont{and} \bibinfo{author}{\bibfnamefont{I.}~\bibnamefont{Lesanovsky}}, \bibinfo{journal}{Phys. Rev. E} \textbf{\bibinfo{volume}{108}}, \bibinfo{pages}{064104} (\bibinfo{year}{2023}{\natexlab{b}}), \urlprefix\url{https://link.aps.org/doi/10.1103/PhysRevE.108.064104}.

\end{thebibliography}

\end{document}